# Programmable Integrated Magnonic Meshes

Piero Florio[1†], Matteo Vitali[1†], Valerio Levati[1], Rasheed M. Ishola[2,3], Luca Ciaccarini Mavilla[4], Nora Lecis[2], Carsten Dubs[5], Riccardo Bertacco[1], Marco Madami[4], Silvia Tacchi[6], Daniela Petti[1]*, Edoardo Albisetti[1]*

[1] Dipartimento di Fisica, Politecnico di Milano, Piazza Leonardo da Vinci 32, 20133 Milano, Italy.
[2] Dipartimento di Meccanica, Politecnico di Milano, via La Masa 1, 20156, Milano, Italy
[3] Dipartimento di Scienze e Tecnologie Aerospaziali, Politecnico di Milano, via La Masa 34, 20156, Milano, Italy
[4] Dipartimento di Fisica e Geologia, Università di Perugia, Via A. Pascoli, Perugia, Italy
[5] INNOVENT e.V., Technologieentwicklung, Jena, Germany
[6] Istituto Officina dei Materiali - Consiglio Nazionale delle Ricerche (CNR-IOM), Sede Secondaria di Perugia c/o Dipartimento di Fisica e Geologia, Università di Perugia, Perugia, Italy

† These authors contributed equally to this work.

*Corresponding authors.
Email: daniela.petti@polimi.it (DP); edoardo.albisetti@polimi.it (EA)

**Integrated circuits have transformed information technology by combining individual functional elements into increasingly complex architectures. Translating this approach to magnonics would enable a highly tunable, compact and energy-efficient solution for wave-based signal processing. However, so far, experimental demonstrations have been mostly limited to isolated elements or passive interconnects, severely limiting the overall functional complexity and scalability.**

**Here we realize multi-stage programmable networks, by monolithically cascading the key elements of magnonic circuitry. Using a single-step direct laser writing process in yttrium iron garnet, we first demonstrate coupled waveguides supporting coherent periodic power transfer over several coupling lengths, and integrated phase shifters providing arbitrary tunable phase delays. We then combine these primitives into reconfigurable devices for splitting, demultiplexing and phase-controlled routing of spin waves. Finally, we realize real-time reconfigurable interferometric meshes for on-chip radio-frequency signal routing, comprising up to nineteen directional couplers and phase shifters in seven cascaded stages, preserving coherent spin-wave propagation over hundreds of wavelengths without intermediate amplification.**

**These direct-write reconfigurable networks bridge a long-standing gap in magnonic scalability, offering a viable pathway toward integrated, large-scale architectures for classical and quantum processing.**

Since the early vision of large-scale integration by Moore[1], the emergence of hardware capable of combining a limited set of functional elements in increasingly complex circuits has marked the beginning of the integrated-circuit era[2], enabling the growth in computing functionalities. More recently, a similar evolution has been pursued in wave-based computing[3–5]. In particular, in photonics, a paradigm shift has taken place, from fixed-function devices towards programmable architectures, where waveguides, phase-shifters and tunable couplers are interconnected into meshes to perform operations based on interference[6–9].

In this context, translating this approach in the field of magnonics would enable compact, reconfigurable information processing directly at microwave frequencies[10–14]. Magnonics exploits spin waves (SWs)[15–18], the collective excitations of magnetic moments in ordered magnetic media, which combine coherent propagation without charge transport, tunability with external stimuli, intrinsic nonlinearity and nanometre-scale wavelengths at microwave frequencies[19,20], enabling ultimate device miniaturization[21]. These features[22] make spin waves appealing for radiofrequency signal processing[23], analogue and neuromorphic computing[24–28] and quantum technologies[29,30].

Yttrium iron garnet (YIG) is the key material platform for magnonics as its exceptionally low magnetic damping enables spin-wave propagation lengths over hundreds of micrometres[31–35]. However, preserving these low-loss properties and phase coherence across devices is extremely challenging, so that cascading two subsequent functional elements requires external amplification[36].

Addressing this critical bottleneck requires a fabrication strategy able to realize monolithically low-loss magnonic devices and interconnects, with high throughput and uniformity for large-scale integration.

Here, we demonstrate that coherent spin-wave circuitry in crystalline YIG thin films can be translated from isolated elements to wafer-scale, programmable, cascaded architectures. Using direct laser writing to locally amorphize crystalline YIG thin films with nanoscale spatial resolution, we define low-loss, magnonic devices within a continuous film. We realize compact phase shifters and tunable directional couplers able to set arbitrary amplitude splitting of spin-wave signals. By integrating these elements through waveguiding structures, we demonstrate phase-controlled 2 x 2 routers able to modulate both power distribution and phase, and large-scale programmable networks for on-chip signal routing, where spin-wave transport and robust phase control are preserved over hundreds of wavelengths across multiple cascaded stages. These results establish integrated magnonics as a viable platform for on-chip wave-based information processing.

## RESULTS

### Laser-written large-scale spin-wave circuits

Spin-wave nanostructures were defined in 100 nm-thick single-crystal Yttrium Iron Garnet (YIG) films via direct irradiation with a focused continuous-wave 405 nm diode laser (see Methods). At this wavelength, the optical penetration depth in YIG exceeds the film thickness, so that the entire volume of the film is irradiated[37] (see Supplementary Note 1). We observed that, above a certain laser fluence threshold, and for irradiation time in the sub-microsecond range, the irradiation triggers a highly localized structural phase-transition from crystalline to amorphous. This is compatible with a laser-induced heating of YIG above the melting threshold, followed by quenching in the amorphous phase due to the rapid cooling down to room-temperature, similarly to what occurs in phase-change materials. While the pristine crystalline YIG is a ferrimagnet supporting low-loss spin-wave propagation, the laser-irradiated amorphous regions become non-magnetic, effectively suppressing spin-wave transport. By exploiting this purely thermal, laser-induced phase nanoengineering[38–42], we directly define high-quality magnonic nanostructures on large areas.

Figure 1a illustrates the concept of a laser-written magnonic mesh network, where spin-wave phase shifters and directional couplers are monolithically interconnected through waveguide structures, to implement photonic-like interferometric transformations.

First, we realized a fundamental building block of magnonic circuitry, namely isolated spin-wave waveguides. The structures were fabricated by irradiating the material surrounding a narrow pristine channel, acting as the guiding core. By adjusting the laser writing parameters and exposure geometry, we defined waveguides with widths ranging from a few micrometers down to the sub-micrometer scale, and controlled, uniform spacing down to < 300 nm.

Direct evidence of the sharp transition between crystalline and laser-induced amorphous regions is provided by Electron BackScatter Diffraction (EBSD) imaging (Extended Data Fig. 1). In the irradiated regions, the EBSD signal is completely suppressed, indicating the loss of long-range crystalline order and the formation of an amorphous phase. By contrast, the waveguides exhibit strong signal arising from the [111] crystallographic orientation of the pristine YIG film.

The magnetic properties of the nanostructures were investigated using Magnetic Force Microscopy (MFM), as shown in Fig. 1b. The amorphized regions exhibit no magnetic contrast, consistent with their non-magnetic character. In contrast, the waveguides, magnetized perpendicular to their axis, show strong magnetic contrast localized at their edges, originating from the closure of stray field lines, confirming the preservation of ferrimagnetism within the crystalline channel, and the low edge-roughness of the structures.

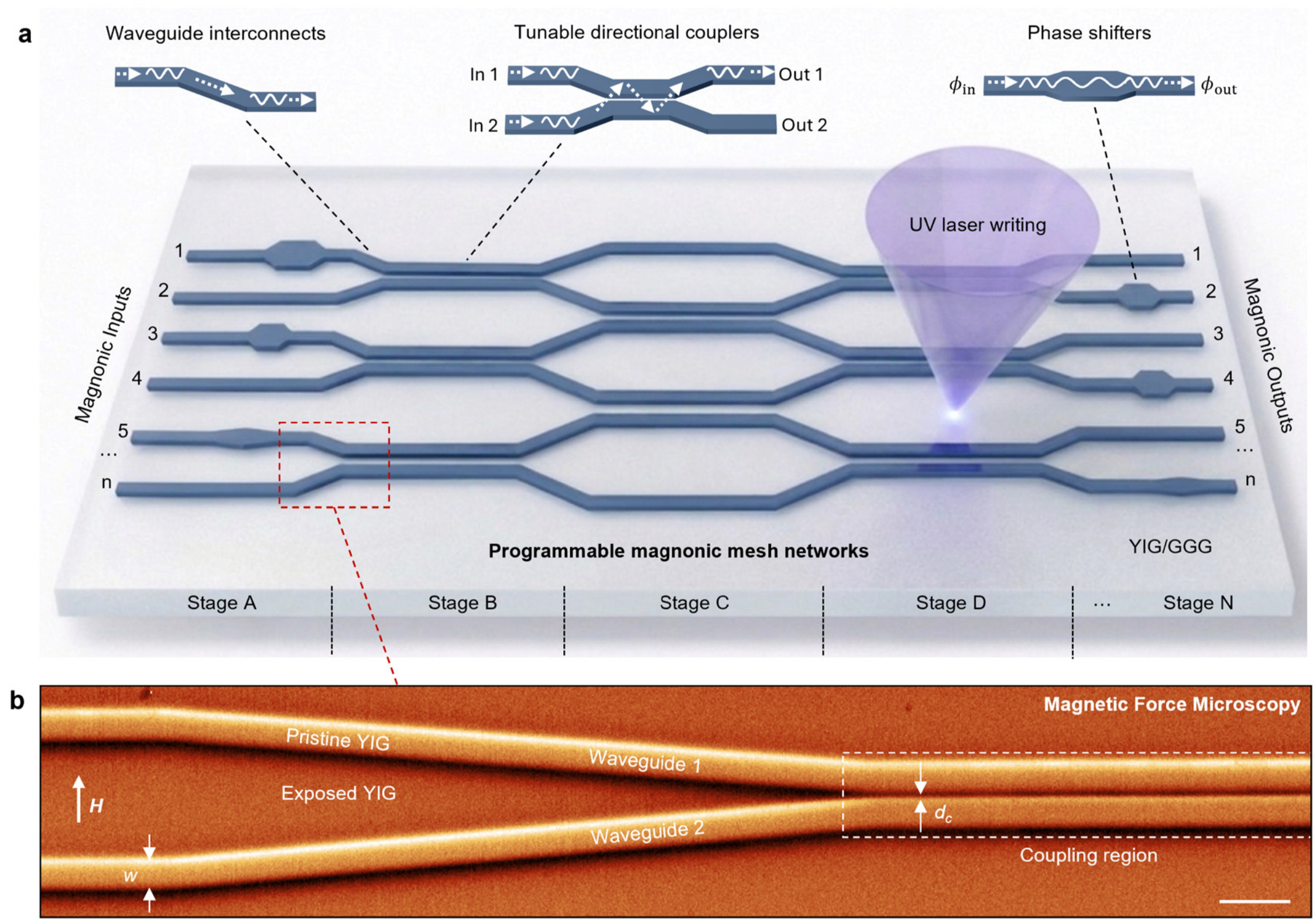

**Figure 1. Laser-written magnonic devices and programmable interferometric meshes.**

**a,** Schematic of the programmable magnonic mesh architecture.The network is fabricated on a crystalline Yttrium Iron Garnet (YIG)/GGG thin film using a single-step, UV laser writing process. Sub-µs UV irradiation locally quenches the YIG into an amorphous, non-magnetic phase, creating nano-magnonic circuits on large scales without the need for additional processing. The monolithically cascaded stages (A through N) consist of directional couplers, phase shifters and waveguide interconnects, allowing for complex, reconfigurable transformations between magnonic inputs and outputs.

**b,** Magnetic Force Microscopy image of a coupling region (right), where two closely spaced parallel waveguides allow spin-wave coupling. The image reveals the contrast between the pristine ferrimagnetic YIG (waveguides) and the laser-exposed non-magnetic regions, upon application of an in-plane magnetic field *H*. Scale bar: 5 µm.

**Long-distance spin-wave propagation and energy transfer in coupled waveguides**

Spin waves, excited by microwave stripline antennas, are studied via Time-Resolved Magneto-Optical Kerr Effect (TR-MOKE) microscopy, which enables two-dimensional phase-resolved mapping with sub-μm resolution (see Methods). In our experimental geometry, an oscillating magnetic field $h_{RF}$ generated by the antenna drives the magnetization dynamics, while a static magnetic field $H$ is applied in-plane perpendicular to the waveguide axis and to the spin-wave wavevector $k$.

Isolated waveguides with different geometries were fabricated and characterized in the 0.9–11.7 GHz frequency range (see Extended Data Fig. 2, 3). Remarkably, a strong spin-wave signal is observed over distances exceeding 650 μm even in sub-μm-wide waveguides - to our knowledge, the longest reported in the literature - highlighting the low magnetic losses introduced by the laser-writing process.

We next realized coupled waveguides, a key element of wave circuits enabling controlled energy transfer between channels[43]. The experimental configuration, in Fig. 2a, comprises two closely spaced parallel waveguides. The antenna excites spin waves exclusively in the top waveguide ("Excited"), while the other remains passive ("Idle").

This system, where the coupling between the two channels occurs via dipolar fields, supports a high-energy symmetric and a low-energy antisymmetric normal mode (See Fig. 2b and Extended Data Fig. 4a)[44–47], which are characterized by different wavevectors, $k_a$ and $k_s$, at fixed excitation frequency $f$. In each waveguide, the propagating spin wave is given by the superposition of these two normal modes. The wavevector mismatch between them, $\Delta k = k_\mathrm{a} - k_\mathrm{s}$ , gives rise to a spatial interference pattern (beating). This results in a periodic modulation of the spin-wave amplitude and, consequently, in a periodic energy exchange between the two waveguides, defining the coupling length $L_\mathrm{C} = \pi/\Delta k$ as the distance required for a complete power transfer (see Supplementary Note 2). At the same time, the fast oscillations are governed by the average wavevector, yielding an effective wavelength of the coupled system $\lambda_C = 4\pi/(k_\mathrm{a} + k_\mathrm{s})$.

We directly observe these spatial beatings experimentally in Fig. 2c, where the profile extracted along the coupled waveguides from the TR-MOKE measurement (Fig. 2d) is reported. In the map, the dark/light contrast corresponds to opposite signs of the out-of-plane component of the magnetization dynamics associated with the spin wave propagation. Initially, all the spin-wave power is in the excited top waveguide. As the wave propagates, the energy periodically oscillates between the two channels with a periodicity of $2L_\mathrm{C}$ that is in agreement with micromagnetic simulations (see Supplementary Methods 1, Extended Data Fig. 4b,c). Crucially for extended circuits, coherent energy transfer and phase coherence persist over several oscillation periods, with no significant additional losses with respect to isolated waveguides.

This low-loss coherent coupling enables magnonic devices that route and redistribute spin-wave signals while remaining compatible with subsequent cascading. Here, we realized a programmable magnonic directional coupler[44] where two initially separate waveguides couple over a finite length $L$ and then separate again (Fig. 2e). In this configuration, only one input arm is excited (input vector [0, 1]), and the output signals are collected in both branches after the coupling region. The output powers and phases delay can be described by a linear transformation of the inputs (see Supplementary Note 2), defined by the transfer matrix of the device:

$$\begin{bmatrix} \mathrm{Out1} \\ \mathrm{Out2} \end{bmatrix} = \mathbf{A}(\vartheta) \begin{bmatrix} 0 \\ 1 \end{bmatrix}$$

This is summarized in Fig. 2f, where the normalized output powers $P_\mathrm{out1}$ and $P_\mathrm{out2}$, and the relative output phase delay $\Delta\phi_\mathrm{out} = \phi_\mathrm{out1} - \phi_\mathrm{out2}$, are calculated as a function of the coupling parameter $\vartheta = \frac{\pi L}{2L_C}$ (see Supplementary Discussion 1 for comparison between experimental and theoretical values). Real-time programmability stems from the fact that the coupling parameter, and therefore the power splitting between the two outputs and their relative phase delay, can be tuned via the external field and excitation frequency.

Experimental validation of this variable splitter functionality is provided by the TR-MOKE maps shown in Fig. 2g–j, acquired at a fixed excitation frequency, at different fields (see Supplementary Methods 2). We demonstrate a continuous tuning of the split ratio, where the input signal is distributed between the two outputs with either a lagging phase ($\pi/2$, Fig. 2g) or a leading phase ($-\pi/2$, Fig. 2i) relative to one another. Alternatively, the power can be almost entirely transferred to either branch (Fig. 2h, "bar state", Fig. 2j "cross state"). The same device can be used as a frequency demultiplexer[48], where two frequency components are spatially separated. To demonstrate this, we excited the same input sequentially with two frequencies, $f_1$ = 3.06 GHz and $f_2$ = 3.15 GHz, at the same external field, observing a nearly complete transfer of the signal from Output 1 (Fig. 2j) to Output 2 (Fig. 2k).

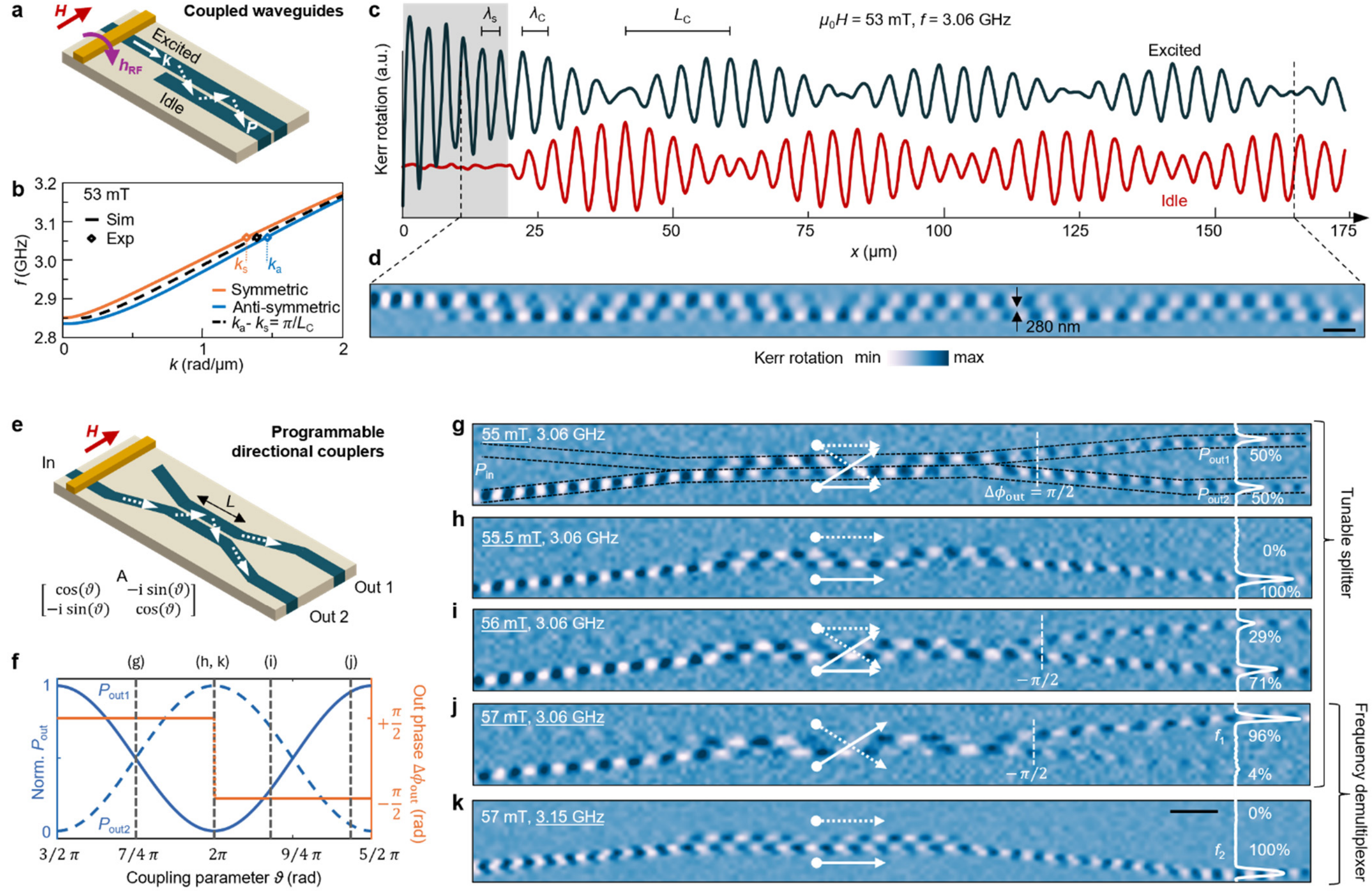

**Figure 2. Low-loss spin-wave coupling and phase control in programmable directional couplers.**

**a,** Schematic of the coupled waveguides geometry, with two 2.2 µm-wide, 200 µm-long waveguides separated by an edge-to-edge gap of 280 nm. Spin waves are initially excited only in the top waveguide.

**b,** Simulated dispersion relations of the coupled system, showing the splitting into symmetric (orange) and antisymmetric (blue) eigenmodes. Points represent experimental data measured at an external field $\mu_0 H = 53$ mT and spin-wave frequency $f$ = 3.06 GHz.

**c, d** Spatial evolution of the spin-wave phase and amplitude along the propagation direction. The TR-MOKE map (d), and extracted line plots (c) demonstrate periodic power transfer between the "Excited" and "Idle" channels over multiple coupling lengths $L_c$. A characteristic $\pi/2$ phase delay is maintained during the energy exchange, a fingerprint of coherent coupling demonstrating the low loss and high homogeneity of the structures. Scale bar: 5 µm.

**e, f** Schematic of the directional coupler (e), where two waveguides are coupled over a finite length $L$, and characteristics of the device (f). The transfer matrix A defines the relationship between the input and outputs, where the coupling parameter $\vartheta$ determines the power distribution $P_{\text{out1,2}}$ and relative phase delay $\Delta\phi_{\text{out}}$. Vertical dashed lines in (f) indicate experimental values extracted from panels (g–k).

**g–j,** Tunable power splitting operation. TR-MOKE measurements at 3.06 GHz show that varying the external magnetic field $H$ tunes the device from a splitter with lagging phase ($\Delta\phi_{\text{out}} = \pi/2$, (g)) or leading phase ($\Delta\phi_{\text{out}} = -\pi/2$, (i)), to a near-complete power transfer to either branch in the "bar state" (h) or "cross state" (j).

**k,** Frequency demultiplexer operation. At a fixed field of 57 mT, changing the excitation frequency to 3.15 GHz reroutes spin waves from the lower branch to the upper branch, allowing signals to be spatially sorted based on their spectral content while preserving phase coherence and low losses. Scale bar: 10 µm.

**Phase shifters and phase-controlled 2x2 routers**

Following waveguides and directional couplers, we now introduce a third key building block: an integrated tunable phase shifter[49,50]. In our platform, this functionality is achieved through a localized widening of the waveguide (Fig. 3a), which modifies the spin-wave wavevector (See Extended Data Fig. 3f), thereby introducing a phase shift $\varphi$, which can be tuned via external field.

Phase shifters with two different geometries are integrated in isolated waveguides. The phase accumulation within the phase shifter is directly visualized by TR-MOKE maps, as shown in Fig. 3b, and can be analyzed by comparing line profiles along the propagation direction before (orange) and after (blue) the phase shifter, for different device geometries and magnetic field conditions in Fig. 3c.

The two phase-shifters yield $-\pi/2$ and $\pi/2$, respectively, at a field of 55 mT, resulting in a relative phase delay of $-\pi$ between the outputs. The same device can be tuned and "switched off" by varying the external magnetic field, as shown by the bottom trace, where a zero net phase shift is achieved at 48 mT. Width-modulated sections emerge as versatile, low-loss on-chip phase shifters that can be engineered through device geometry and tuned via external fields.

Waveguides, tunable couplers and phase shifters are the fundamental building blocks for programmable wave-based circuitry. Here, we combine these three elements to realize a 2 x 2 device for routing spin waves based on their relative input phase delay, namely a reconfigurable phase-controlled router, a key device in wave-based signal processing networks[51,52].

To do this, a phase-shifter on the top input branch is cascaded with a directional coupler (Fig. 3d). For simplicity, both arms are excited with in-phase spin waves with the same input power (input vector [1, 1]). We fabricated devices with two different phase shifter geometries, resulting in a phase shift of $\varphi_1 \simeq -\pi/2$ (Fig. 3f) and $\varphi_2 \simeq \pi/2$ (Fig. 3i) at 53 mT.

The overall device functionality is captured in Fig. 3e, where the output power in the first branch $P_{\text{out1}}$ (contour lines) and the relative phase delay between the outputs $\Delta\phi_{\text{out}}$ (color map) are calculated as a function of the coupling parameter $\vartheta$ and the phase shift $\varphi$. Since the two input arms are initially excited in phase, $\varphi$ also corresponds to the relative input phase delay at the entrance of the directional coupler:

$$\begin{bmatrix} \text{Out1} \\ \text{Out2} \end{bmatrix} = \mathbf{A}(\varphi)\mathbf{B}(\vartheta) \begin{bmatrix} 1 \\ 1 \end{bmatrix}$$

Experimental data points (stars), represent distinct device states obtained by varying the external field, visualized in the TR-MOKE maps below (see Supplementary Discussion 2 for comparison between experimental and theoretical values).

Fig. 3f-h demonstrate the phase-controlled routing mechanism, where the functionality of the device is programmed via external magnetic fields. Spin waves excited in input 1 and 2 are initially in phase ($\Delta\phi_{\text{in}} = 0$). Then, the phase shifter imposes a phase shift of $\varphi_2 \simeq \pi/2$ , creating a relative input phase delay, that is maintained until the entrance of the directional coupler. Here, the signals are coupled and routed towards the top branch, at 53 mT, or bottom branch at 55 mT. At an intermediate field of 53.5 mT (Fig. 3g), where the phase shifter imposes a different $\varphi_2 \simeq \pi$, the antisymmetric mode of the coupler is excited, so that the signal output is perfectly balanced in the two branches. The opposite scenario is shown in Fig. 3i,j, where a different phase shifter geometry yields $\varphi_1 \simeq -\pi/2$, giving rise to the opposite routing behavior and field dependence. Importantly, a strong signal output is retained, allowing these devices to be further cascaded without the need for intermediate amplification stages.

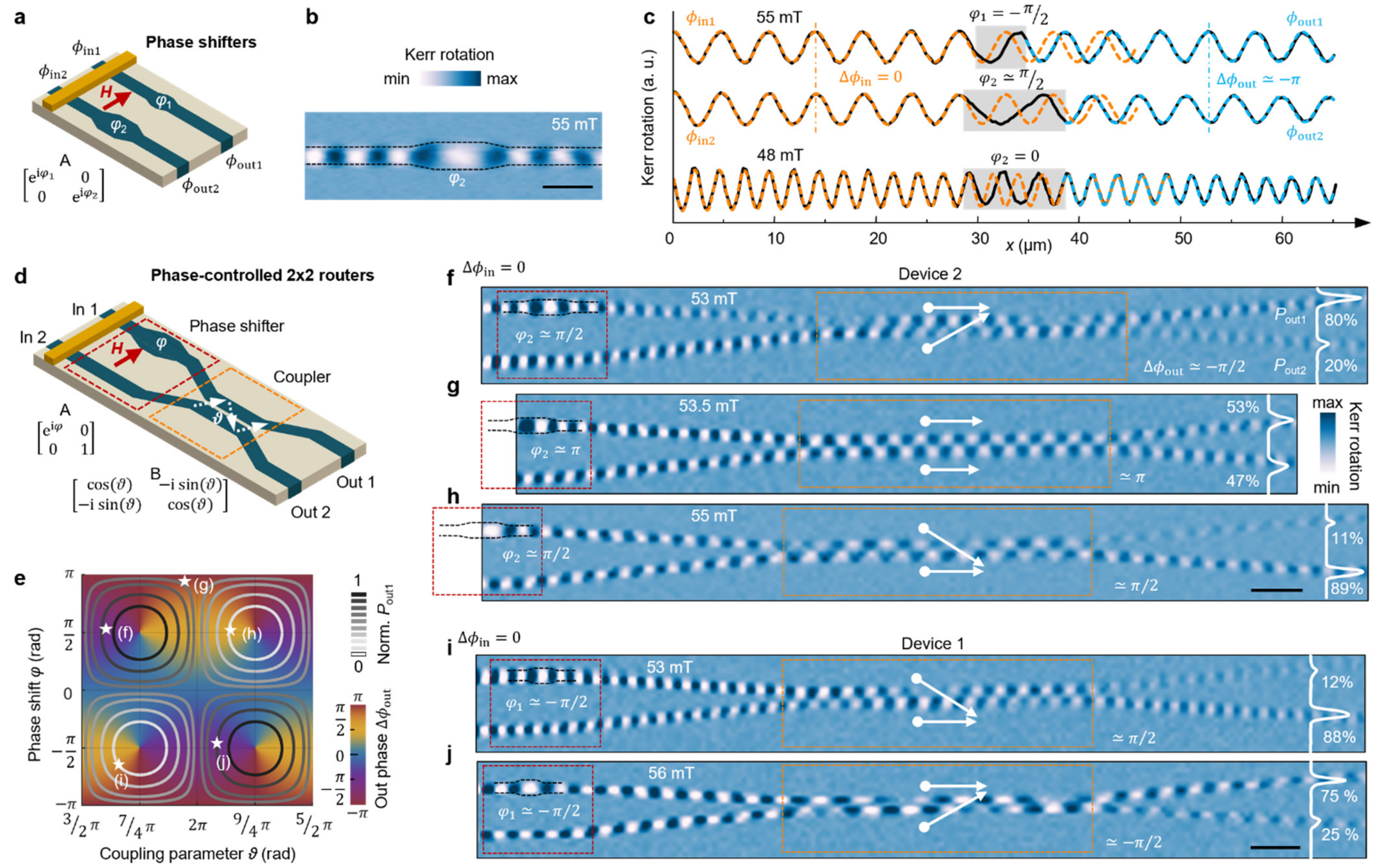

**Figure 3. Spin-wave phase shifters and programmable 2 x 2 phase-controlled routers.**

**a,** Schematic of the phase-shifting element. A localized widening of the waveguide modifies the spin-wave wavevector, resulting in a net phase shift $\varphi$ relative to the nominal waveguide.

**b, c** Experimental validation via TR-MOKE. The spatial map (b) and corresponding line profiles (c) illustrate phase accumulation along the widened section for different geometries and external fields. At a fixed field of 55 mT, different geometries yield phase shifts of $\varphi_1 = -\pi/2$ (top trace) and $\varphi_2 = \pi/2$ (middle), resulting in a total relative phase delay of $\Delta\phi_{\text{out}} = -\pi$ between the two signals. Field-tunability is shown in the bottom trace, where the same device at 48 mT yields a zero net phase shift, demonstrating reconfigurable operation. Scale bar: 5 µm.

**d,** Schematic of a programmable 2×2 phase-controlled router, where the input signals are combined and routed according to their relative phase delay. It consists of a phase shifter A cascaded with a directional coupler B.

**e,** Characteristics of the device, mapping the normalized output power $P_{\text{out1}}$ (contours) and relative phase delay $\Delta\phi_{\text{out}}$ (color map) as a function of phase shift $\varphi$ and coupling parameter $\vartheta$. Stars indicate experimental values from panels (f-j).

**f–j** TR-MOKE maps show phase-controlled routing in two distinct devices. The initially in-phase signals are combined and routed towards the top (f, j), bottom (h, i), or equally split between the two branches (g), by tuning the phase shift $\varphi$ and coupling parameter $\vartheta$ via device geometry and reconfiguring them in real-time via external fields. Scale bars: 10 µm.

## Programmable magnonic mesh networks

We finally demonstrate scaled-up spin-wave mesh circuits by cascading multiple elements into large-scale mesh networks, in which output states result from successive coherent phase and amplitude transformations across multiple coupling stages. The device layout is shown in Fig. 4a and comprises 4 input channels, 4 outputs and 4 cascaded stages with 5 coupling nodes and 2 phase shifters. Phase shifters are implemented on inputs 2 and 4 (stage A), followed by a sequence of coupling regions: first between channels 1–2 and 3–4 (stage B), then between 2–3 (stage C), and finally again between 1–2 and 3–4 (stage D). This design ensures that each input can influence each output, and that each input signal can eventually be routed towards any output arm.

The propagation of spin waves through the entire mesh is directly visualized by TR-MOKE microscopy maps. Figures 4b–d show representative maps acquired at an external field of 53 mT at increasing distances from the excitation antenna, where the different functional stages can be clearly identified. The measured phase and amplitude patterns reveal coherent transport and coupling across the successive coupling regions. Remarkably, in Fig. 4e-g, we show that the mesh state can be reconfigured in real-time to route the input radio-frequency signals towards all the possible combinations of the four outputs between balanced output power (e), single output (f), two outputs (g) and three outputs (h) (see Supplementary Methods 2). In Extended Data Fig. 5, we demonstrate operation of the mesh in a wide frequency range, preserving programmability of the routing functionality from 1.35 GHz up to 10.08 GHz. Further mesh complexity is demonstrated in Extended Data Fig. 6, where a 6-input, 6-output mesh with 7 cascaded stages and 16 coupling nodes is realized, where spin-wave signal at 3.06 GHz is detected more than 150 wavelengths and 700 μm away from the excitation, with a spin-wave decay length across the whole device network of 123 μm and an attenuation of 0.07 dB/μm. These cascaded device networks overcome the high losses and poor phase control typically associated with magnonic devices, which are critical bottlenecks preventing scalability towards chip-scale architectures.

## CONCLUSIONS

In summary, we demonstrate that coherent spin-wave circuitry can be scaled from individual elements to large, cascaded and programmable architectures. To do so, we use a single-step direct laser writing approach that operates in ambient conditions with high throughput (> 1.8 $cm^2$/h), without etching, additional material processing, or charged beams. We realize the fundamental building blocks of integrated magnonic circuitry, namely phase shifters and tunable couplers, interconnected through waveguiding structures, and combine them into programmable splitters, frequency demultiplexers and phase-controlled routers. We then cascade these elements, realizing programmable magnonic interferometric meshes for on-chip radio-frequency signal routing, maintaining coherent transport across multiple stages without intermediate amplification.

Future integration of active and passive elements for local tuning of the network, such as localized magnetic fields[53], electrostatic gates[54], thermal effects[55] or mechanical actuators[56] could allow node-addressable operation.

More broadly, these results unlock a path towards tailored large-scale magnonic chips for signal processing, linear and non-linear wave-based computing and quantum technologies.

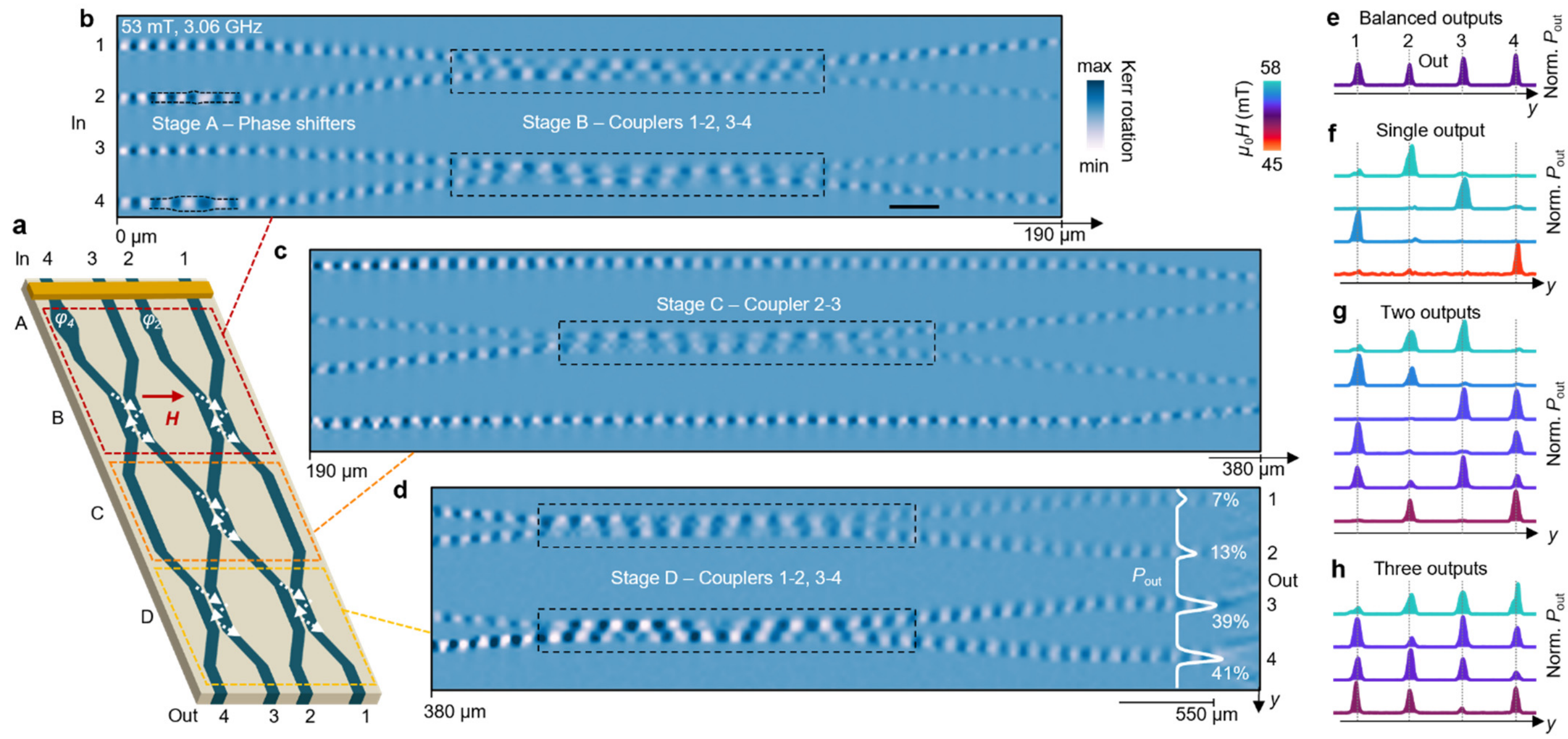

**Figure 4. Programmable magnonic mesh networks.**

**a,** Schematic of a four-input, four-output mesh network device. The design integrates phase shifters (Stage A) followed by three cascaded coupling stages (Stages B–D). This interconnected architecture allows each input signal to influence each output.

**b-d,** TR-MOKE maps visualize coherent spin-wave propagation and routing across the mesh, at increasing distances from the excitation antenna. Robust signal and preserved phase coherence are observed after the phase shifters, multiple coupling stages and bent waveguide interconnects, over distances exceeding 500 µm from the antenna and > 130 wavelengths.

**e-h,** Normalized output power at 3.06 GHz, demonstrating the network programmability. By tuning the external magnetic field to control phase and coupling, the input signals can be routed to produce all the possible configurations of balanced power across all outputs (e), routing to a single output (f), power splitting across two or three outputs (g, h).

Increased mesh complexity, with 6 inputs and 6 outputs, 16 coupling nodes in 7 cascaded stages is shown in Extended Data Figure 6. Scale bars: 10 µm.

## METHODS

### Samples and antennas fabrication

The 100 nm-thick single-crystal YIG film employed in this study was grown via liquid phase epitaxy (LPE) on a 500 µm-thick GGG(111) substrate. To excite spin waves, stripline antennas with widths ranging from 1.3 to 2 µm were fabricated using maskless photolithography, followed by thermal deposition of Cr(10 nm)/Au(100 nm) and subsequent lift-off.

### Direct Laser Writing

Laser patterning of the YIG thin films was performed using a commercial maskless laser writing system (Heidelberg Instruments DWL 66+) equipped with a continuous-wave 405 nm semiconductor diode laser. The system operates in High-Resolution Mode, providing a minimum feature size below 300 nm. The laser beam is automatically focused onto the sample surface, and the sample is scanned using a high-speed stage at a fixed velocity of 1800 mm $s^{-1}$. Patterns are defined on a 50×50 $nm^2$ pixel grid, allowing precise spatial control over the laser exposure. In this configuration, a throughput of up to ~1.8 $cm^2$/h is achieved, enabling rapid fabrication of complex magnonic structures across large-scale areas. As an example, the exposure time of the mesh networks in Fig. 4 and Extended Data Fig. 6 was lower than 1 second in both cases.

### Magnetic Force Microscopy

MFM measurements were performed using a Park Systems NX10 atomic force microscope operating in lift mode. PPP-LM-MFMR magnetic probes from Nanosensors were used for magnetic contrast imaging. During measurement, an external magnetic field of 50 mT was applied using a Caylar Magnetic Field Module system integrated with the microscope. Topographic information was simultaneously acquired during each scan. The resulting data were processed to remove polynomial background trends and to correct for scan line distortions and tip-related artifacts.

### Electron Backscattered Diffraction

Inverse Pole Figure maps from electron backscatter diffraction (EBSD) were collected using a Zeiss Sigma 500 field-emission scanning electron microscope, fitted with an Oxford C-NANO EBSD detector.

### TR-MOKE measurements

Spin waves are detected using a stroboscopic TR-MOKE setup. A microwave antenna, patterned on the sample by conventional lithography, is driven by a radio-frequency (RF) signal at 12 dBm, generating an oscillating Oersted field that excites spin waves at the driving frequency. To probe the spin dynamics, we exploit the magneto-optical Kerr effect (MOKE). A train of polarized laser pulses is focused normally onto the sample surface, and the polarization rotation of the reflected beam, which is proportional to the out-of-plane (OOP) component of the magnetization, is recorded as a function of position. The optical source is a picosecond diode laser, Picoquant LDH-IB-640-B or LDH-IB-405-B, emitting at 638 nm and 403 nm respectively, with a pulse duration below 100 ps. This pulse width defines the temporal resolution of the experiment, limiting the detectable spin-wave frequencies to below 12 GHz. The RF excitation and the laser pulse train are synchronized providing a stroboscopic snapshot of the dynamics. The measurement is performed point-by-point by scanning the sample position with a piezoelectric stage. The laser spot size is below 1 µm, defining the spatial resolution. A 2D map (or a 1D line scan) of the spin-wave amplitude and phase is reconstructed by compiling the MOKE signal at each spatial coordinate. The resulting maps are interpolated using bicubic interpolation to improve visual continuity, and each map is normalized to its own maximum amplitude.

### Micromagnetic Simulations

Micromagnetic simulations were performed by numerically solving the Landau–Lifshitz-Gilbert equation using the GPU-accelerated software Mumax3[57]. The detailed procedures are reported in Supplementary Methods 1.

## DATA AVAILABILITY

The data that support the findings of this study are available in the Zenodo repository with identifier DOI:10.xxxx/zenodo.xxxxxxx

## ACKNOWLEDGMENTS

This work was partially performed at PoliFab, the micro- and nanotechnology center of the Politecnico di Milano. The authors thank the PoliFab staff, particularly Stefano Bigoni for assistance with wire bonding, Giacomo Corrielli and Francesco Morichetti for discussions on integrated photonics, and Christian Back and Christian Riedel for their support in setting up the TR-MOKE experimental system, Giorgio Ferrari for components for the TR-MOKE setup. E.A. acknowledges funding from the European Union's Horizon 2020 research and innovation programme under grant agreement number 948225 (project B3YOND) and from the FARE programme of the Italian Ministry for University and Research (MUR) under grant agreement R20FC3PX8R (project NAMASTE). E.A. and S.T. acknowledge funding from the European Union – Next Generation EU – "PNRR – M4C2, investimento 1.1 – "Fondo PRIN 2022" – TEEPHANY– ThreEE-dimensional Processing tecHnique of mAgNetic crYstals for magnonics and nanomagnetism ID 2022P4485M CUP D53D23001400001". D.P. acknowledges funding from the European Union – Next Generation EU – "PNRR – M4C2, investimento 1.1 – "Fondo PRIN 2022" – PATH – Patterning of Antiferromagnets for THz operation id 2022ZRLA8F – CUP D53D23002490006" and from Fondazione Cariplo and Fondazione CDP, grant n° 2022-1882. R.B. acknowledges funding from NextGenerationEU, PNRR MUR – M4C2 – Investimento 3.1, project IR_0000015 – "Nano Foundries and Fine Analysis – Digital Infrastructure (NFFA–DI)", CUP B53C22004310006. C.D. acknowledges funding from the Deutsche Forschungsgemeinschaft (DFG, German Research Foundation)-271741898.

## AUTHOR CONTRIBUTIONS

P.F., M.V. and V.L performed the laser nanopatterning. M.V., V.L., L.C.M., M.M., and S.T. performed the magnetic characterisation. M.V. performed the micromagnetic simulations. V.L., R.M.I., and N.L. performed the structural characterization. P.F. fabricated the spin-wave antennas and performed the TR-MOKE measurements. C.D. provided the YIG samples. P.F., M.V. and V.L contributed to data analysis. R.B. contributed to the development of the TR-MOKE setup. P.F., M.V., V.L., D.P and E.A. wrote the manuscript with contributions from all the authors. All authors contributed to discussion and data interpretation. D.P. and E.A. designed and supervised the research.

## COMPETING INTERESTS

## ADDITIONAL INFORMATION

Supplementary Information is available for this paper.

Correspondence and requests for materials should be addressed to Daniela Petti and Edoardo Albisetti.

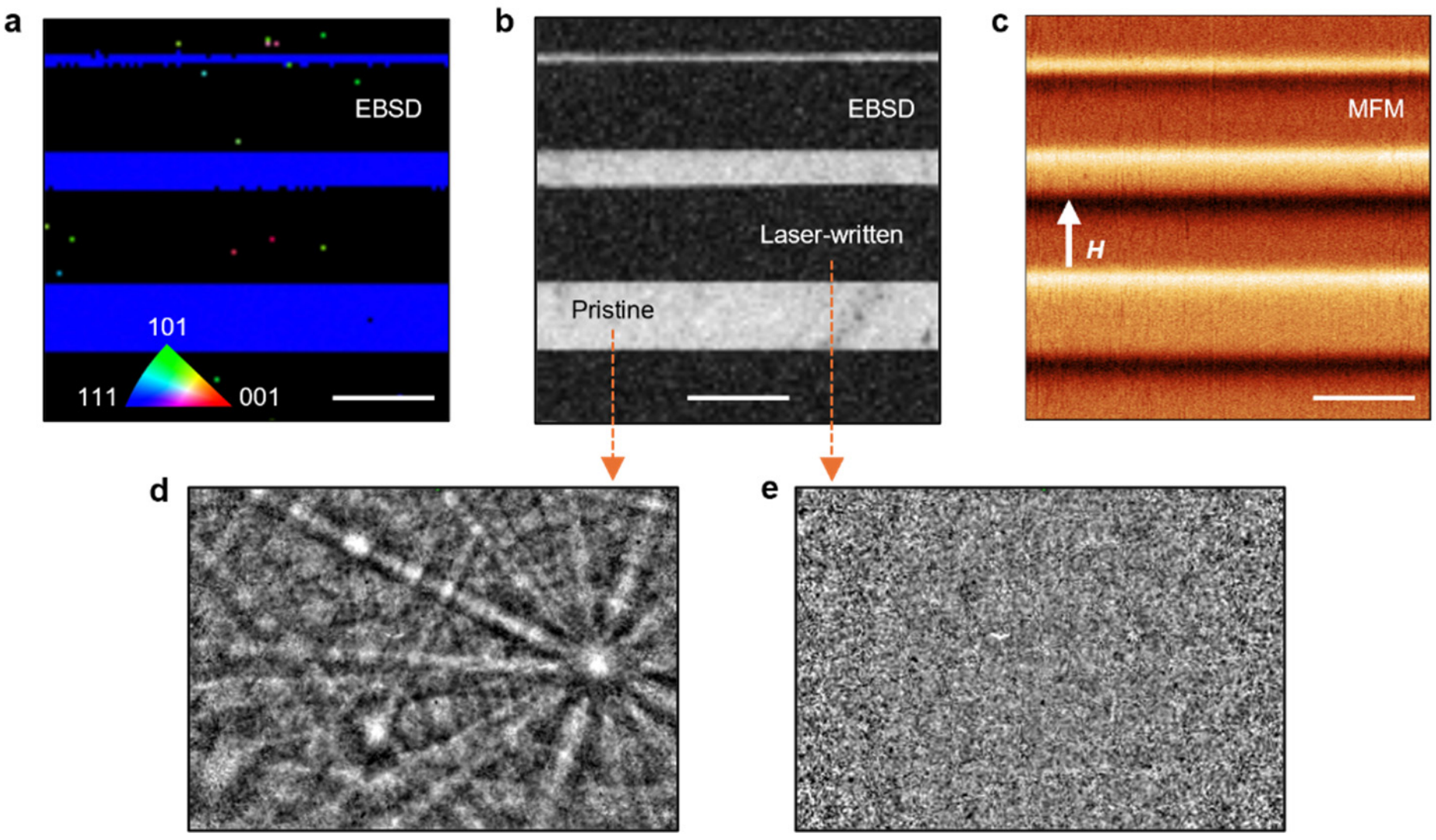

**Extended Data Figure 1. Structural and magnetic characterization of laser-written waveguides.**

**a,** Inverse Pole Figure (IPF-Z) obtained via Electron BackScatter (EBSD) Diffraction mapping of parallel waveguides down to sub-µm width, showing the unexposed pristine conduits in blue, and the exposed amorphous region, in black, featuring a sharp transition from crystalline to amorphous phase.

**b**, EBSD band contrast map corresponding to (a). Bright horizontal bands indicate the crystalline waveguides, separated by dark regions associated with amorphized material that does not produce a detectable diffraction signal.

**c**, Magnetic Force Microscopy image of isolated waveguides magnetized by an in-plane magnetic field *H*. The magnetic contrast localized at the waveguide edges confirms the ferrimagnetic nature of the pristine crystalline YIG and the absence of magnetic signal from the surrounding non-magnetic amorphous regions.

**d,e,** Kikuchi patterns acquired via EBSD from a pristine crystalline waveguide (d) and from a laser-written amorphous region (e), showing well-defined diffraction bands in the former and the absence of any recognizable pattern in the latter, consistent with the loss of long-range crystalline order.

Scale bars: 5 µm.

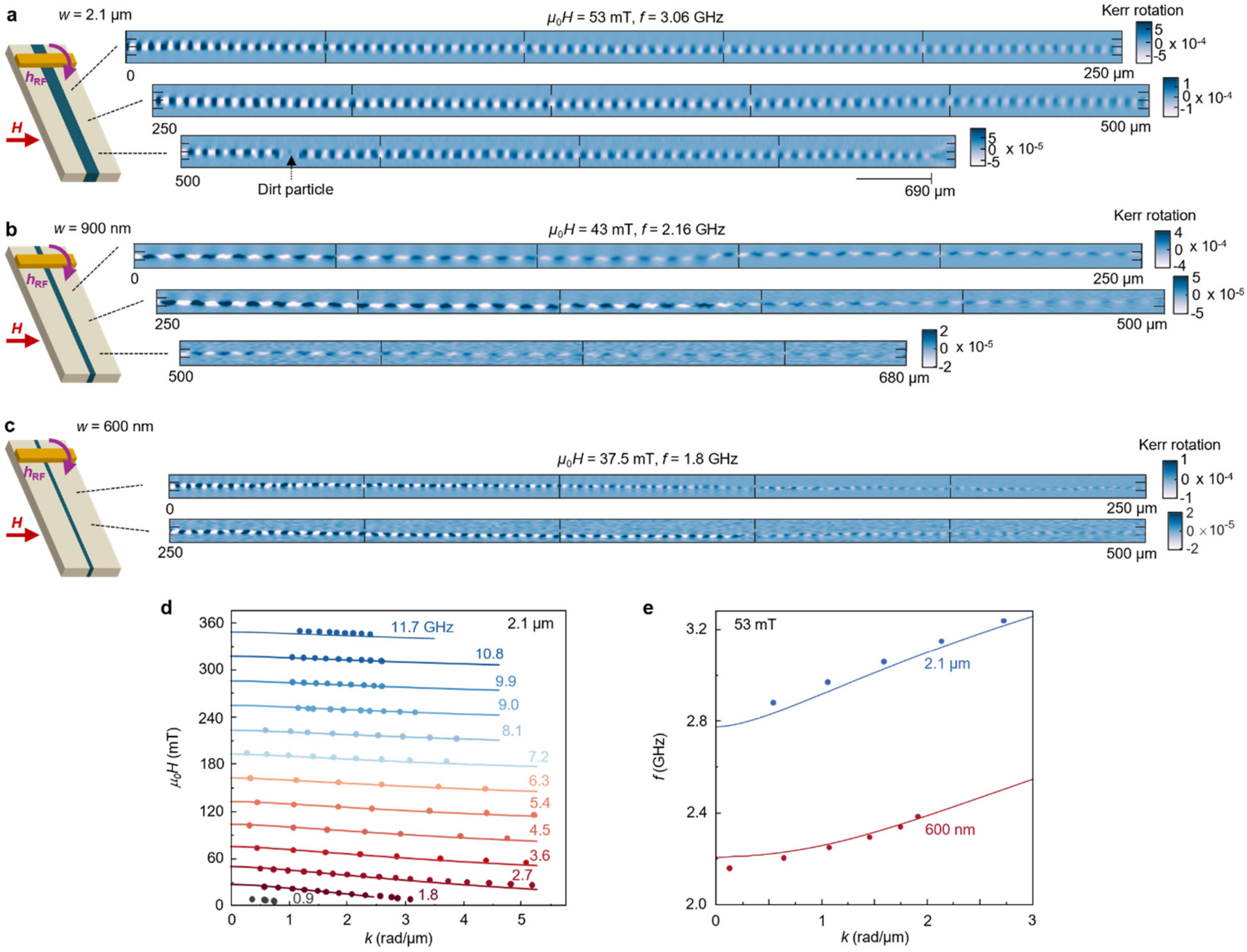

**Extended Data Figure 2. Long-distance propagation and dispersion of isolated laser-written waveguides.**

**a-c,** Time-Resolved Magneto-Optical Kerr Effect (TR-MOKE) microscopy maps of spin waves propagating in straight waveguides with 2.1 μm (a), 900 nm (b), and 600 nm (c) widths, acquired at increasing distances from the excitation region, with the external magnetic field and frequency decreasing from (a) to (c). Coherent spin-wave propagation is observed over remarkably long distances, up to nearly 700 μm in (a) and (b), and 500 μm in (c).

**d,** Magnetic-field dependence of the wavevector $k$ for the fundamental $n = 1$ mode in a 2.1 μm-wide waveguide (markers), compared with simulations (curves), showing the tunability of the waveguide dispersion across a broad frequency range from 0.9 GHz to 11.7 GHz.

**e,** Experimental (markers) and simulated (curves) frequency–wavevector dispersion relations for 600 nm and 2.1 μm-wide waveguides at $\mu_0 H = 53$ mT. The dispersion of the narrower waveguide is downward shifted relative to that of the wider waveguide, due to the reduced effective magnetic field, showing the effects of the waveguide geometry over the spin waves dispersion.

The details of the simulations are reported in Supplementary Methods 1.

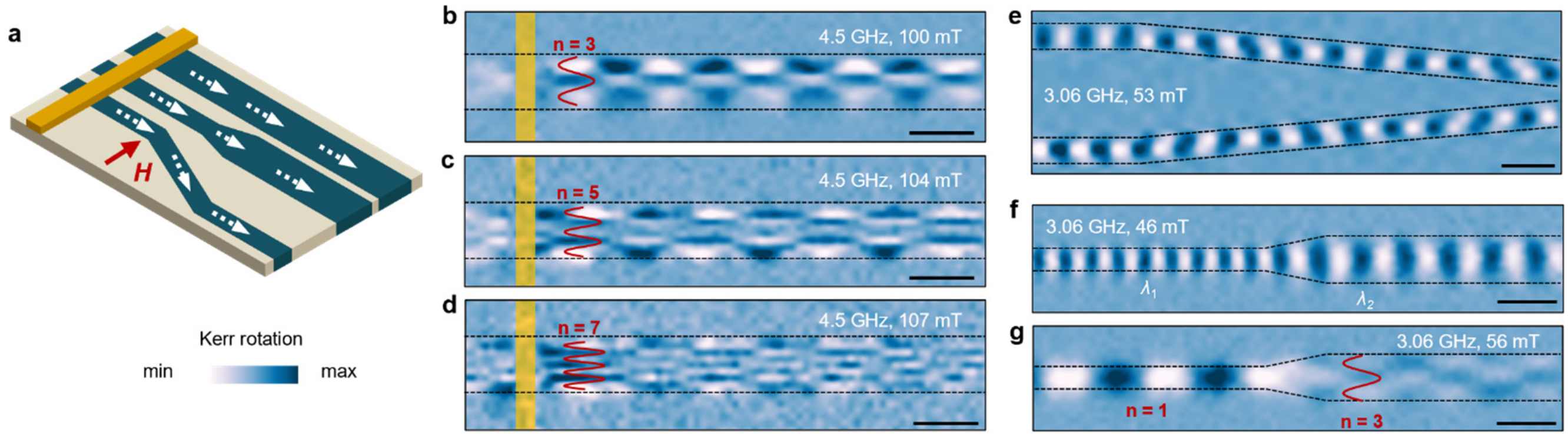

**Extended Data Figure 3. Higher order guided modes, curved and width-modulated waveguides.**

**a,** Schematic of various waveguide geometries, including straight, bent, and variable-width waveguides.

**b-d,** TR-MOKE maps of spin waves propagating in a 4.3 µm-wide waveguide, showing transverse standing modes of order $n = 3$ (a), $n = 5$ (b) and $n = 7$ (c) across the waveguide width. All three maps were acquired at a fixed frequency of 4.5 GHz while increasing the external magnetic field from (a) to (c). The shaded yellow regions indicate the position of the excitation antenna.

**e,** TR-MOKE map of spin waves in 2 µm-wide bent waveguides with spin waves maintaining their coherence during the propagation.

**f-g,** TR-MOKE maps of spin waves propagating in a waveguide with variable width, showing modulation of the spin wave wavelength (f) and mode conversion (g). In (g) the fundamental $n = 1$ mode excited in the narrower 2.25 µm-wide section evolves into a higher-order $n = 3$ mode in the wider 4.25 µm region of the waveguide. Even-order modes are not excited, as the driving field has a spatial profile symmetric with respect to the waveguide axis, yielding a non-zero overlap only with odd-order transverse modes.

In all panels the external magnetic field is applied perpendicular to the spin-wave propagation direction (Damon–Eshbach configuration).

Scale bars: 5 µm.

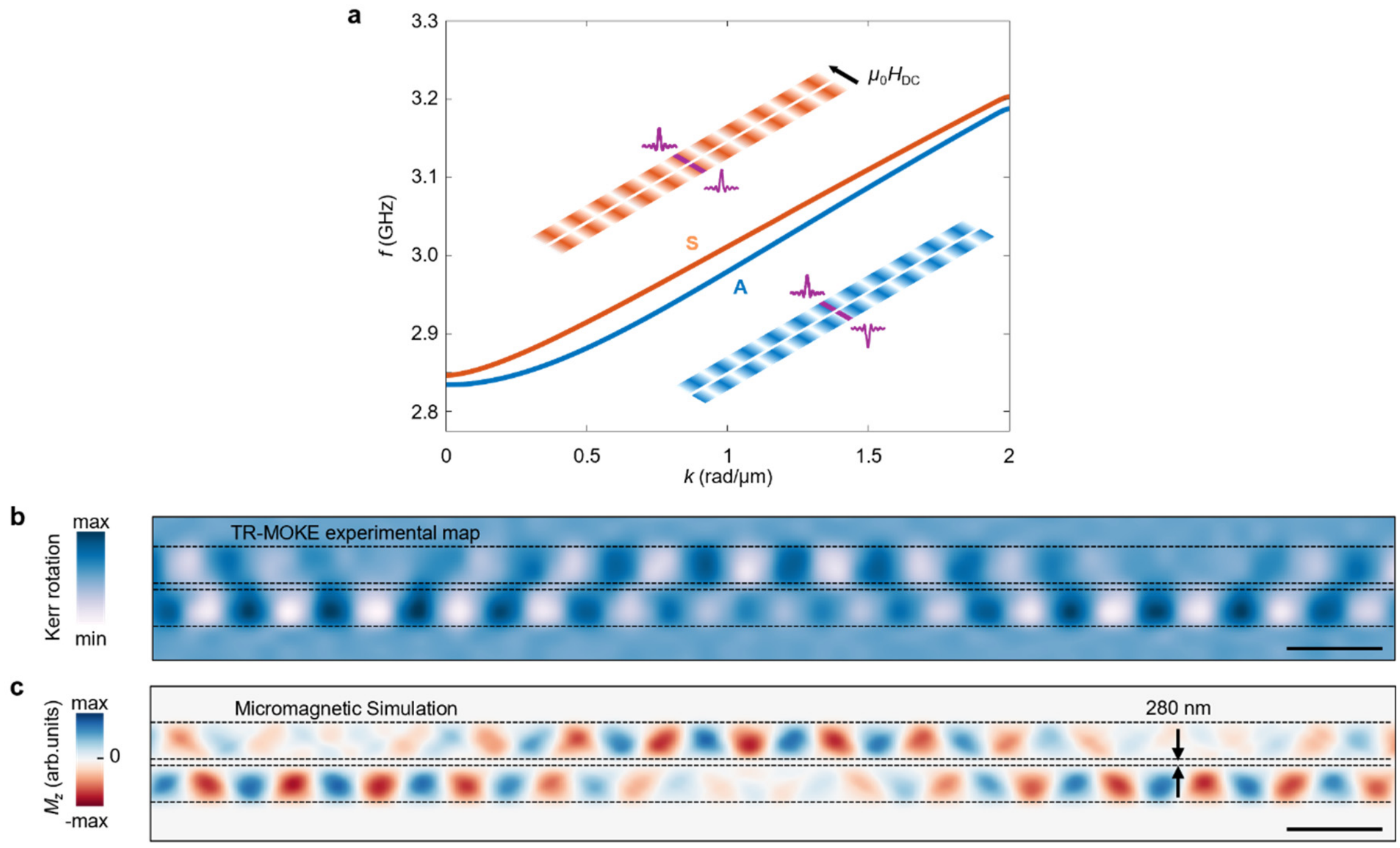

**Extended Data Figure 4. Dispersion curves and spin-wave propagation in coupled waveguides.**

**a,** Dispersion relations of the symmetric (orange line) and antisymmetric (blue line) modes, obtained from two separate simulations by directly exciting the symmetric (top) and antisymmetric (bottom) normal modes (see Supplementary Methods 1). The plot demonstrates that the symmetric mode is upwards-shifted compared to the antisymmetric one. Insets depict the qualitative spin-wave spatial configurations and their respective excitations (in violet) for both modes at a fixed wavevector.

**b,c,** Zoomed view of the TR-MOKE map (b) of the coupled waveguides in Fig. 2d and corresponding micromagnetic simulation (c). Scale bars: 5 µm.

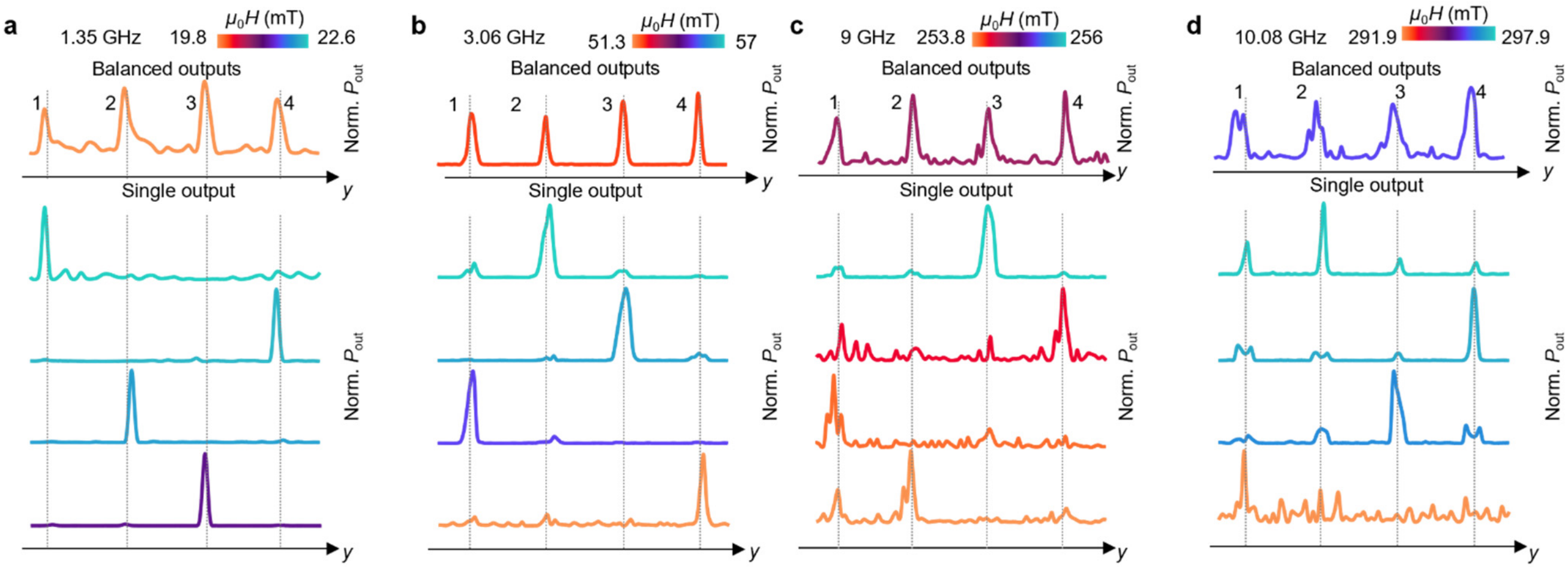

**Extended Data Figure 5. Frequency-Dependent Operation of the Magnonic Mesh Network.**

**a-d,** Output powers of the mesh network shown in Fig. 4 measured across different frequencies while varying the external magnetic field. Significant spin-wave signals are observed, along with programmable routing of the input signal either to individual outputs or to multiple outputs simultaneously, within the frequency range of 1.35 GHz to 10.08 GHz.

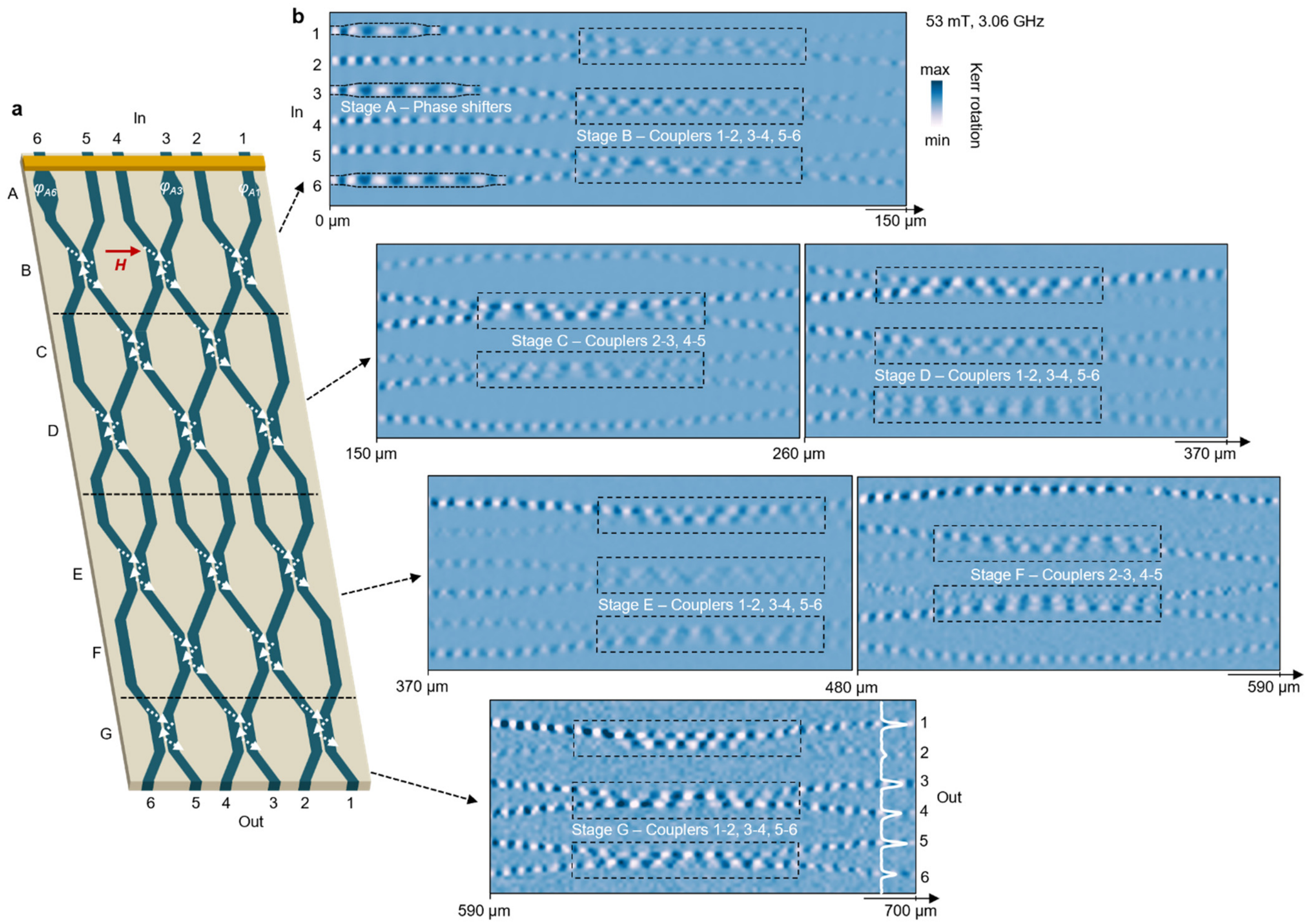

**Extended Data Figure 6. 6-input, 6-output, 7 stages magnonic mesh network**

**a,** Schematic of the mesh network. Phase shifters (stage A), integrated at inputs 1, 3 and 6, are cascaded with six subsequent coupling stages (B-G).

**b,** TR-MOKE maps of all stages of a 6-input, 6-output magnonic mesh comprising 16 coupling nodes. Spin waves propagate and couple within the network and a strong signal is detected at the outputs more than 150 wavelengths and 700 µm away from the inputs, with a decay length of 123 µm and attenuation of 0.07 dB/µm.

# SUPPLEMENTARY INFORMATION FOR:

## Programmable Integrated Magnonic Meshes

Piero Florio[1†] , Matteo Vitali[1†], Valerio Levati[1], Rasheed M. Ishola[2,3], Luca Ciaccarini Mavilla[4], Nora Lecis[2], Carsten Dubs[5], Riccardo Bertacco[1], Marco Madami[4], Silvia Tacchi[6], Daniela Petti[1]*, Edoardo Albisetti[1]*

[1] Dipartimento di Fisica, Politecnico di Milano, Piazza Leonardo da Vinci 32, 20133 Milano, Italy.
[2] Dipartimento di Meccanica, Politecnico di Milano, via La Masa 1, 20156, Milano, Italy
[3] Dipartimento di Scienze e Tecnologie Aerospaziali, Politecnico di Milano, via La Masa 34, 20156, Milano, Italy
[4] Dipartimento di Fisica e Geologia, Università di Perugia, Via A. Pascoli, Perugia, Italy
[5] INNOVENT e.V., Technologieentwicklung, Jena, Germany
[6] Istituto Officina dei Materiali - Consiglio Nazionale delle Ricerche (CNR-IOM), Sede Secondaria di Perugia c/o Dipartimento di Fisica e Geologia, Università di Perugia, Perugia, Italy

† These authors contributed equally to this work.

*Corresponding authors.
Email: daniela.petti@polimi.it (DP); edoardo.albisetti@polimi.it (EA)

## TABLE OF CONTENTS

**Supplementary Methods 1:** Details on micromagnetic simulations.

Micromagnetic simulations were performed using the software MuMax3[1]. For all simulations, the following micromagnetic parameters for YIG were used: saturation magnetization $M_s$ = 140 kA/m, exchange stiffness $A$ = 4 pJ/m, and Gilbert damping α = $10^{-4}$.

### Field vs wavevector maps for isolated waveguide

The simulations were performed to compute field–wavevector ($\mu_0 H_{\mathrm{DC}}$–$k$) maps, in which each value of the external saturating field $\mu_0 H_{\mathrm{DC}}$ is associated with the wavevector $k$ of a single spin-wave mode excited at a fixed frequency and propagating in a waveguide of width $w$ = 2.1 μm. The simulated waveguide had dimensions of 200 μm × 2.1 μm × 100 nm along the $x$, $y$ and $z$ axes respectively, discretized into cells of 250 nm × 30 nm × 100 nm. A schematic of the geometry is represented in Fig. S1a.

The system was initially saturated along the $y$-direction, i.e., along the width of the waveguide, by a magnetic field $\mu_0 H_{\mathrm{DC}}$. Spin waves were excited by an oscillating magnetic field applied within a 1 μm-wide rectangular region defined by $|x - x_{\mathrm{c}}| \leq$ 0.5 μm, where the excitation center $x_{\mathrm{c}}$ is located 30 μm from the negative $x$-boundary of the structure. The excitation field was spatially modulated along the $y$-direction using a cosine profile to selectively excite the $n = 1$ standing mode of the waveguide:

$$\mu_0 \mathbf{H}_{\mathrm{AC}} = \mu_0 H_0 \cos\left(\frac{\pi y}{w}\right) \sin(2\pi f t)\, \hat{\mathbf{x}}$$

where $\mu_0 H_0$ = 0.3 mT, $w$ = 2.1 μm, and $f$ is the excitation frequency. The simulation runtime was set to 50 ns.

A one-dimensional fast Fourier transform (1D-FFT) along the $x$-direction was performed on the final spatial map of the out-of-plane magnetization component $M_{\mathrm{z}}$, restricting the analysis to a region where spin waves propagate with a $k > 0$ wavevector . The magnitude of the FFT was then averaged over all $y$-cells, yielding the dominant propagation wavevector in the waveguide.

The simulation was repeated for multiple values of the external field $\mu_0 H_{\mathrm{DC}}$, increased in steps of 0.4 mT. This procedure provided the corresponding wavevector for each field value. The resulting data were organized into a colormap with $\mu_0 H_{\mathrm{DC}}$, and $k$ as axes, and the color scale representing the squared FFT magnitude.

The dispersion curves shown in Extended Data Fig. 2d were extracted directly from this colormap. The entire procedure was repeated for different excitation frequencies $f$, each explored over a specific field range.

### Dispersion relation of isolated waveguides

The simulations were carried out to calculate the dispersion relation of spin waves propagating in Damon-Eshbach configuration in waveguides with widths of 2.1 μm and 600 nm. The simulated geometry, presented in Fig. S1a, had dimensions of 200 μm × 2.1 μm or 200 μm × 600 nm along the $x$ and $y$ axes respectively, with a thickness of 100 nm along the $z$ axis. The computational grid was discretized into cells of 100 nm × 30 nm × 100 nm.

The system was initially saturated by applying a uniform external magnetic field $\mu_0 H_{\mathrm{DC}}$ = 52.2 mT along the $y$-direction. The small discrepancy regarding the magnetic field compared to the experiments can be attributed to the uncertainty in measuring the actual applied field. Spin waves were excited by an impulse field applied within a rectangular region at the center of the waveguide defined by $|x| \leq$ 0.75 μm. The excitation field is given by:

$$\mu_0 \mathbf{H}_{\mathrm{AC}} = \mu_0 H_0 \cos\left(\frac{\pi y}{w}\right) \mathrm{sinc}(2\pi f (t - t_0))\, \hat{\mathbf{x}}$$

where $\mu_0 H_0$ = 5 mT, $f$ = 10 GHz, and $t_0$ = 5 ns, while $w$ denotes the width of the waveguide. The cosine term provides spatial modulation along the $y$-direction, enabling efficient excitation of the symmetric $n$ = 1 standing mode across the waveguide width.

The total simulation time was set to 200 ns, and magnetization maps were recorded with a temporal resolution of 0.1 ns. To suppress spin-wave reflections, the damping parameter was gradually increased up to $\alpha = 1$ near the edges along the $x$-direction.

The dispersion relation was obtained by averaging along the $y$-direction the magnitude of the two-dimensional fast Fourier transform (2D-FFT), computed over time and the $x$-coordinate, of the dynamic out-of-plane magnetization component: $M'_{\mathrm{z}}(t,x,y) = M_{\mathrm{z}}(t,x,y) - M_{0,\mathrm{z}}(0,x,y)$, where $M_0$ represents the initial saturated ground state. The dispersion curves shown in Extended Data Fig. 2e were extracted from the resulting 2D-FFT power colormap.

**Dispersion relation of coupled waveguides**

To obtain the dispersion relation of the coupled waveguides shown in Fig. 2b, the structure composed of two YIG waveguides in Fig. S1b was considered. The overall dimensions of the simulated geometry were 400 µm × 4.68 µm × 100 nm along the $x$, $y$ and $z$ axes, discretized into cells of 100 nm × 20 nm × 100 nm. Each waveguide has a width of $w$ = 2.2 µm and the two are separated by a 280 nm gap of non-magnetic material (modelled as empty space).

The system was initially saturated along the $y$-direction, by a uniform magnetic field $\mu_0 H_{\mathrm{DC}}$ = 52.2 mT. Spin waves were excited by an impulse field applied within a rectangular region ($|\, x \,| \leq 0.75$ µm) located at the center of the top waveguide. The excitation field is defined as:

$$\mu_0 \mathbf{H}_{\mathrm{AC}} = \mu_0 H_0 \cos\left(\frac{\pi(y - y_{\mathrm{c}})}{w}\right) \mathrm{sinc}\big(2\pi f(t - t_0)\big)\, \hat{\mathbf{x}}$$

where $\mu_0 H_0$ = 5 mT, $f$ = 10 GHz, and $t_0$ = 5 ns. Here, $w$ and $y_c$ denote the width and the $y$-coordinate of the center of the top waveguide, respectively. The cosine term, spatially restricted to the top waveguide, ensures efficient excitation of the fundamental $n$ = 1 mode across its width.

To suppress spin-wave reflections, the damping parameter was gradually increased up to $\alpha$ = 1 near the edges along the $x$-direction. The total simulation time was set to 200 ns, and magnetization maps were recorded with a temporal resolution of 0.1 ns.

The simulation data were processed using the same procedure adopted for the isolated waveguides. In particular, the dispersion relation was extracted from the power colormap obtained via a two-dimensional fast Fourier transform (2D-FFT).

**Dispersion relation of symmetric and antisymmetric modes in coupled waveguides**

The simulations described in this section were conducted to identify the symmetric and antisymmetric modes within the dispersion relation of the coupled waveguides of Fig. 2b in the Damon-Eshbach configuration. Symmetric (antisymmetric) modes are characterized by an in-phase (out of phase) precession of the magnetization in the two dipolarly coupled waveguides (see Extended Data Fig. 4a).

The micromagnetic simulation setup closely follows the procedure outlined in the section describing the dispersion relation in coupled waveguides, with minor adjustments to the runtime (now set to 400 ns), overall geometry dimensions (800 µm × 4.68 µm × 100 nm along the $x$, $y$ and $z$ axes), and discretization mesh (250 nm × 40 nm × 100 nm). The geometry is sketched in Figure S1b.

The dynamic excitation of spin waves was designed to selectively excite either the symmetric or the antisymmetric mode in the two waveguides simultaneously. To achieve this, both waveguides were excited by two pulses, applied within a rectangular region defined by $|\, x \,| \leq 0.75$ µm. Each pulse has a cosine profile along y, centered in $\pm y_c$ and was defined only over the $y$-range of the corresponding waveguide:

$$\mu_0 \mathbf{H}_{\mathrm{AC,top}} = \mu_0 H_0 \cos\left(\frac{\pi(y - y_{\mathrm{c}})}{w}\right) \mathrm{sinc}\big(2\pi f(t - t_0)\big)\, \hat{\mathbf{x}}$$

$$\mu_0 \mathbf{H}_{\text{AC,bottom}} = \pm\, \mu_0 H_0 \cos\left(\frac{\pi(y + y_c)}{w}\right) \text{sinc}\big(2\pi f(t - t_0)\big)\, \hat{\mathbf{x}}$$

To excite the symmetric (antisymmetric) mode, the pulses were applied with the same (opposite) sign. The simulation data were processed using the same procedure adopted for the dispersion of coupled waveguides, with the key difference that only one branch of the dispersion is obtained depending on the symmetry of the excitation. The dispersions of both symmetric and antisymmetric modes (obtained from two independent simulations) are presented in Extended Data Fig. 4a, highlighting the fact that the antisymmetric mode has a lower frequency for a given value of the wavevector with respect to the symmetric one. This result for the Damon-Eshbach geometry contrasts with the backward volume configuration, where the frequency ordering of the symmetric and antisymmetric modes is known from the literature[2] to be reversed. For Damon-Eshbach modes this result can be understood in terms of dynamic dipolar fields, which are minimized in the antisymmetric mode, while for backward volume the symmetric oscillation favors dipolar flux closure across the gap, shifting the mode to lower frequency.

**2D map of spin wave propagation in coupled waveguides**

The simulation was designed to reproduce the two-dimensional map obtained from time-resolved magneto-optical Kerr effect (TR-MOKE) measurements of spin waves propagating in coupled waveguides (Extended Data Fig. 4c). The simulated geometry had dimensions of 260 µm × 8 µm × 100 nm along the $x$, $y$ and $z$ axes, discretized into cells of 100 nm × 20 nm × 100 nm.

The coupled waveguides were modelled as two waveguides, each having a $w$ = 2.2 µm and separated by a 280 nm gap, positioned symmetrically with respect to the center of the simulated region along the $y$-direction, such that the center of the gap coincides with the center of the simulation area. The top waveguide extends over the entire $x$-range, while the bottom waveguide begins 40 µm from the negative $x$ boundary of the geometry, as in Fig. S1c. The non-magnetic region was modeled as empty space.

The system was initially saturated by a static magnetic field $\mu_0 H_{\text{DC}}$ = 52.2 mT applied along the y-direction. Spin waves were excited exclusively in the top waveguide by an oscillating magnetic field applied within a rectangular region defined by $| x - x_{\text{c}} | \leq 0.75$ µm, where $x_{\text{c}}$ is located 20 µm from the negative $x$-boundary of the waveguide.

The excitation field was spatially restricted to the top waveguide and modulated along the $y$ - direction using a cosine profile. This ensures efficient excitation of the fundamental $n$ = 1 standing mode across its width, and is given by:

$$\mu_0 \mathbf{H}_{AC} = \mu_0 H_0 \cos\left(\frac{\pi(y - y_c)}{w}\right) \sin(2\pi f t)\, \hat{\mathbf{x}}$$

where $\mu_0 H_0$ = 0.2 mT and $f$ = 3.06 GHz. Here, $w$ and $y_c$ represent the width and the $y$-coordinate of the center of the top waveguide, respectively.

To suppress spin-wave reflections, the Gilbert damping parameter was gradually increased up to $\alpha = 1$ near the edges: along both $x$-boundaries for the top waveguide and toward the positive $x$-boundary for the bottom waveguide. The total simulation time was set to $T$ = 200 ns.

The color scale represents the out-of-plane component of the magnetization, defined as: $M'_{\text{z}}(t, x, y) = M_{\text{z}}(t, x, y) - M_{0,\text{z}}(0, x, y)$ , where $M_0$ denotes the initial saturated ground state. The data were further smoothed using a two-dimensional Gaussian filter with a standard deviation of 100 nm along both the $x$- and $y$-directions.

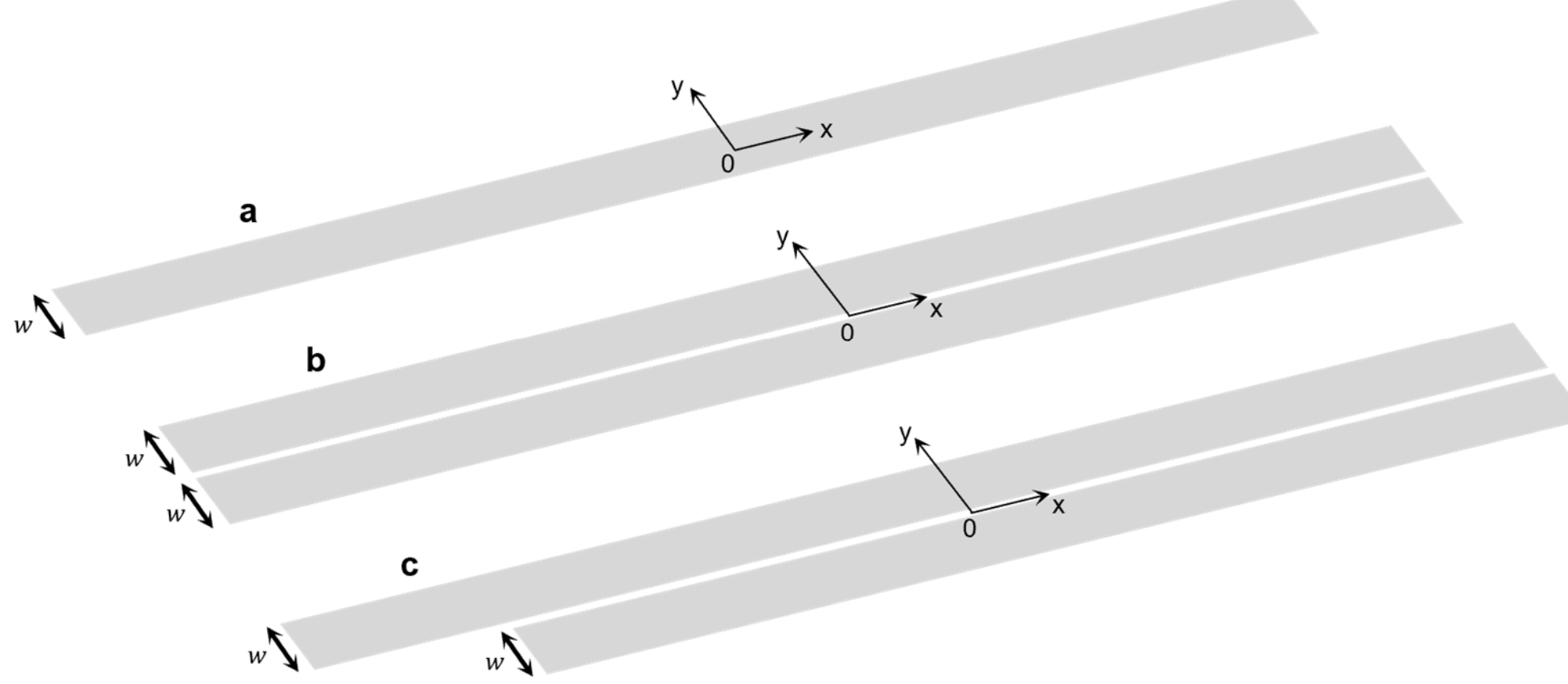

**Figure S1. Geometry of the simulated systems.**
**a-c** Schematic illustrations of the micromagnetic simulation geometries for a single (isolated) waveguide (a) and for two coupled waveguides (b,c).

**Supplementary Methods 2:** Extraction of experimental parameters of directional couplers and mesh network.

**Output power of directional couplers**

Power extraction is performed via a custom MATLAB script. A rectangular region encompassing the two output waveguides downstream of the coupling region is selected from a 2D TR-MOKE map. For each line parallel to the x-axis within this region, the variance is computed, yielding a function that associates a variance value to each y-coordinate and thus reconstructs a profile proportional to the output power. The resulting plot is normalized such that the sum of the areas under the two peaks (corresponding to the two waveguides) equals unity. A peak is considered to end when its tails decay to 5% of its maximum value. The script outputs the normalized profile along with the areas under the two peaks, which are proportional to the percentage of power carried by each waveguide.

**Extraction of the experimental coupling parameter $\boldsymbol{\vartheta_{exp}}$**

The coupling parameter $\vartheta$ is defined as $\vartheta = \frac{\Delta k}{2} L = \frac{\pi}{2} \frac{L}{L_c}$. As shown in the formula, it depends solely on the physical coupling variables between the waveguides, such as the length of the coupling region $L$ and $\Delta k$, which arises from the interaction and is directly related to the distance between the waveguides. As seen in the theoretical model, both normal modes are expected to coexist in the coupled region, leading to a spatial variation of $M_z$ of the type:

$$\sin(k_a x) + \sin(k_s x) = 2 \cos\left(\frac{\Delta k}{2} x\right) \sin\left(\frac{k_a + k_s}{2} x\right)$$

The software fits this function to the spatial $M_z$ profile within the coupled region, thereby extracting $\Delta k$ and consequently $\vartheta$. To facilitate fit convergence, a 1D FFT is first performed to provide approximate estimates of $k_a$ and $k_s$, which are then used as initial fit parameters. This strategy significantly improves the accuracy of $\vartheta$ compared to direct FFT results, as the limited spatial extent of the region of interest prevents the FFT from yielding highly precise values. The fit imposes an additional model constraint, allowing refinement of the wavevector values extracted from the FFT results.

**Input and output phase of directional couplers**

The input or output phase is extracted using a MATLAB-based routine. Two segments are selected within the waveguides of interest on a 2D TR-MOKE map, defined by their start and end points. To ensure consistency, the software constrains both segments to span the same range of $x$-coordinates where x is the propagation direction. For each segment, the $M_z$ profile is extracted and fitted with a sinusoidal function. The experimental phase is then obtained as the difference between the two fitted phases.

**Attenuation losses in the mesh network**

A 2D map obtained from TR-MOKE measurements is loaded by a MATLAB-based script. For a fixed x-coordinate, corresponding to a specific position along the propagation direction of the waveguides, the signal amplitude profile is extracted as a function of the y coordinate, i.e., along the direction perpendicular to the waveguides. On this profile, four points are selected corresponding to the amplitudes of the four waveguides present in the device, and their sum is calculated. This procedure is repeated for each consecutive map, each acquired at increasing x positions along the propagation direction, allowing the reconstruction of the total amplitude (sum of the four channels) as a function of the propagation distance. The obtained values are then fitted with a decreasing exponential model, from which the decay coefficient λ, the decay length, and the propagation losses in dB/μm are extracted.

**Supplementary Note 1:** Laser intensity profile within the YIG film.

The aim of this model is to estimate the optical intensity of the laser within the YIG volume during the patterning process. A model for calculating the intensity profile of an optical radiation propagating in a multilayer system can be found in Ref.[3]. Adapting this model to our system, consisting of a bilayer composed of a YIG layer of thickness $d$ = 100 nm and a GGG substrate, the intensity profile along the $z$ axis in YIG can be obtained as:

$$I_{\mathrm{Y}} = I_0(x(t),y)\frac{n_{\mathrm{Y}}t_{\mathrm{AY}}^2}{\left|1+r_{\mathrm{AY}}r_{\mathrm{YG}}e^{-\alpha d}e^{i\frac{4\pi n_{\mathrm{Y}}}{\lambda}d}\right|^2}\left[e^{-\alpha z}+(r_{\mathrm{YG}})^2e^{-\alpha(2d-z)}+2r_{\mathrm{YG}}e^{-\alpha d}\cos\left(4\pi\frac{n_{\mathrm{Y}}}{\lambda}(d-z)\right)\right]$$

The real and imaginary parts of the YIG refractive index were obtained from the complex dielectric function in Ref.[4] at the laser wavelength $\lambda = 405$ nm yielding $n_Y = 2.806$, and $\kappa = 0.187$. The YIG absorption coefficient was then calculated as $\alpha = 4\pi\kappa/\lambda \simeq 58.023 \times 10^3$ cm$^{-1}$. For the GGG substrate, the refractive index was taken from Ref.[5] as $n_{\mathrm{G}} = 2.015$, while the imaginary part was neglected since GGG is essentially transparent at this wavelength.

Several approximations were adopted to tailor the model to our system. Since $\kappa \ll n_{\mathrm{Y}}$, only the real part of the YIG refractive index was used to calculate the normal-incidence Fresnel coefficients, while the refractive index of air was considered equal to 1:

$$r_{\mathrm{AY}} = \frac{1 - n_{\mathrm{Y}}}{1 + n_{\mathrm{Y}}}, \quad r_{\mathrm{YG}} = \frac{n_{\mathrm{Y}} - n_{\mathrm{G}}}{n_{\mathrm{Y}} + n_{\mathrm{G}}}, \quad t_{\mathrm{AY}} = \frac{2}{1 + n_{\mathrm{Y}}}$$

To account for the moving laser source, the input intensity $I_0(x(t),y)$ was assumed to have a Gaussian profile in the $(x,y)$ plane, translating along the $x$ axis with a speed of $v = 1800$ mm/s:

$$I_0(x(t),y) = \frac{2P_{\mathrm{L}}}{\pi w^2}e^{-2\frac{(x-(x_0+vt))^2+y^2}{w^2}}$$

where $P_L$ is the incident laser power, $x_0$ is the initial position of the laser along the $x$ axis, and $w$ is the laser beam waist radius, assumed independent of the propagation direction $z$ due to the small thickness of the YIG film. Overall, the heat source in the YIG volume is proportional to the intensity and can be expressed as $Q_{\mathrm{Y}} = \alpha I_{\mathrm{Y}}$. The resulting intensity profiles are shown in Fig. S2. The laser scan is assumed to start at $x_0 = -6$ µm. The intensity is evaluated at the point $x = y = z = 0$ as a function of time in Fig. S2a, and as a function of time and depth below $x = y = 0$ in Fig. S2b. The laser power and spot size were set to $P_{\mathrm{L}}$ = 90 mW and $2w$ = 500 nm, respectively.

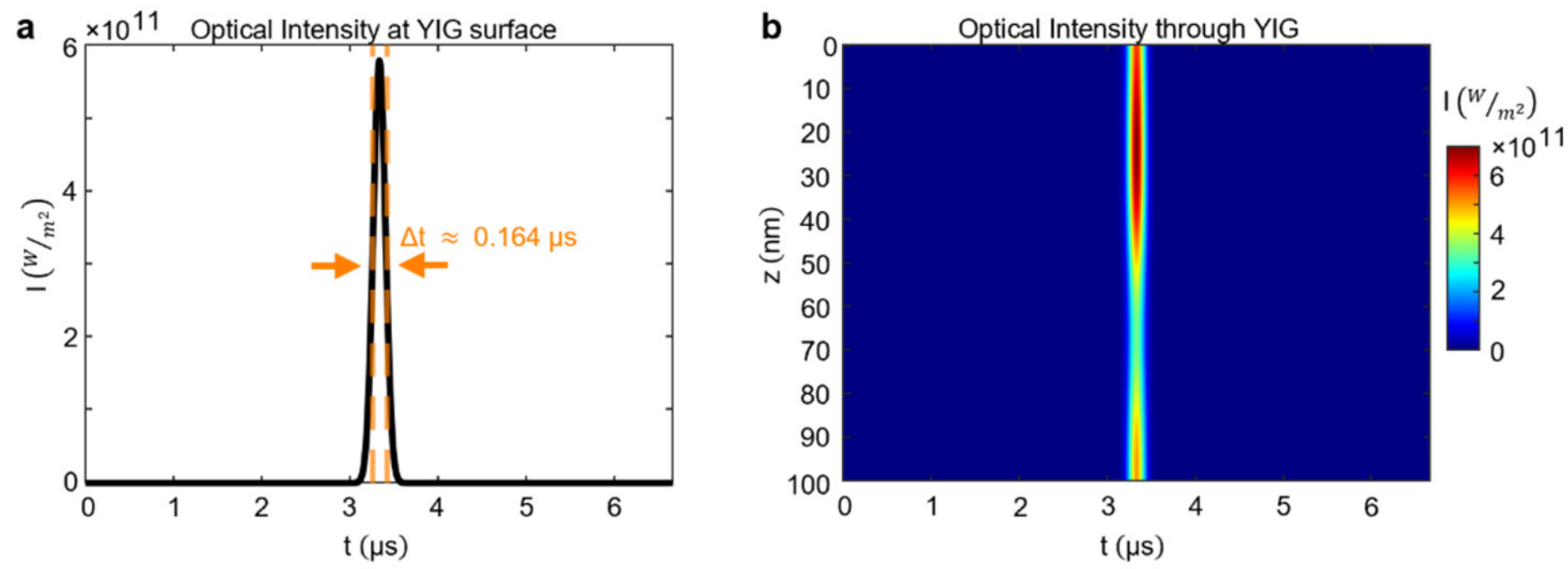

**Figure S2. Laser intensity profile in YIG.**
**a,** Temporal evolution of the laser intensity on the YIG surface at point $(x = y = z = 0)$. The calculated FWHM of $\Delta t \simeq 0.164$ µs highlights the sub-microsecond duration of the intense optical exposure. **b,** 2D colormap showing the optical intensity distribution as a function of time and depth below coordinates $x = y = 0$. Thin-film interference effects modulate the intensity profile along the z-axis, sustaining high energy deposition across the thickness.

**Supplementary Note 2:** Matrix formalism for two-inputs, two-stage devices.

**Directional coupler**

An ideal directional coupler, comprising two identical, lossless waveguides, supports two normal propagating modes: the symmetric mode and the antisymmetric mode. These two modes are eigenstates of the coupled system: each propagates undisturbed, accumulating only a phase along the propagation direction. Denoting the respective wavenumbers as $k_s$ and $k_a$, the evolution of each mode is described by the exponential phase factors $e^{ik_s z}$ and $e^{ik_a z}$.

In this modal basis, the transfer matrix of the system is therefore diagonal.

$$U_{\mathrm{m}}(L) = \begin{bmatrix} e^{ik_s L} & 0 \\ 0 & e^{ik_a L} \end{bmatrix}$$

and the two normal modes can be represented by column vectors of the type:

$$v_s = \begin{bmatrix} 1 \\ 0 \end{bmatrix} \qquad v_a = \begin{bmatrix} 0 \\ 1 \end{bmatrix}$$

Any state can be expanded in the normal-mode basis. However, in this representation, the components of the column vectors do not directly correspond to the amplitudes and phases of the field in the two physical waveguides. Therefore, to obtain a description where each vector component fully represents the field in its respective waveguide, a change of basis is required, transforming from the normal-mode basis to the physical basis of the waveguides.

The change-of-basis matrix M is constructed by arranging, as its columns, the eigenvectors representing the normal modes, expressed in the new physical basis that will be employed from this point onward

$$M = \frac{1}{\sqrt{2}} \begin{bmatrix} 1 & 1 \\ 1 & -1 \end{bmatrix}$$

In the symmetric mode, indeed, the two waveguides are characterized by equal amplitude and equal phase, whereas in the antisymmetric mode there is a π phase shift between the two waveguides, represented by the minus sign. In the waveguide basis, the propagation matrix is obtained as:

$$U_{\mathrm{g}}(L) = M\, U_{\mathrm{m}}(L)\, M^{-1} = \frac{1}{\sqrt{2}} \begin{bmatrix} 1 & 1 \\ 1 & -1 \end{bmatrix} \begin{bmatrix} e^{ik_s L} & 0 \\ 0 & e^{ik_a L} \end{bmatrix} \frac{1}{\sqrt{2}} \begin{bmatrix} 1 & 1 \\ 1 & -1 \end{bmatrix} =$$

$$= \frac{1}{2} \begin{bmatrix} e^{ik_s L} + e^{ik_a L} & e^{ik_s L} - e^{ik_a L} \\ e^{ik_s L} - e^{ik_a L} & e^{ik_s L} + e^{ik_a L} \end{bmatrix}$$

By applying prosthaphaeresis formulas

$$e^{ik_s L} + e^{ik_a L} = 2e^{i\frac{k_s + k_a}{2}L} \cos\left(\frac{k_s - k_a}{2} L\right) = 2e^{i\bar{k}L} \cos\left(\frac{\Delta k}{2} L\right)$$

$$e^{ik_s L} - e^{ik_a L} = 2ie^{i\frac{k_s + k_a}{2}L} \sin\left(\frac{k_s - k_a}{2} L\right) = -2ie^{i\bar{k}L} \sin\left(\frac{\Delta k}{2} L\right)$$

and by suitably defining the arithmetic mean and the difference between the wave numbers of the antisymmetric and symmetric modes, as well as the coupling length:

$$\bar{k} = \frac{k_s + k_a}{2}, \qquad \Delta k = k_a - k_s, \qquad L_c = \frac{\pi}{\Delta k};$$

it is possible to rewrite the matrix representing the directional coupler in the new reference frame:

$$U_{\mathrm{g}}(L) = e^{i\bar{k}L} \begin{bmatrix} \cos\left(\frac{\Delta k}{2} L\right) & -i\sin\left(\frac{\Delta k}{2} L\right) \\ -i\sin\left(\frac{\Delta k}{2} L\right) & \cos\left(\frac{\Delta k}{2} L\right) \end{bmatrix} = e^{i\bar{k}L} \begin{bmatrix} \cos\left(\frac{\pi L}{2L_c}\right) & -i\sin\left(\frac{\pi L}{2L_c}\right) \\ -i\sin\left(\frac{\pi L}{2L_c}\right) & \cos\left(\frac{\pi L}{2L_c}\right) \end{bmatrix}$$

For DE modes in horizontally coupled waveguides, the antisymmetric mode has a lower energy than the symmetric mode, as shown in Extended Data Fig.4a. This implies $k_a > k_s$ at a fixed excitation frequency, which is why $\Delta k$ has been defined as $\Delta k = k_{\mathrm{a}} - k_s$. For convenience, from now on the argument of the trigonometric functions (which is intrinsically positive) will be denoted as $\vartheta$. This parameter intrinsically depends on the geometric properties of the directional coupler and on the external magnetic field: indeed, $L$ represents the physical length over which the waveguides are coupled, whereas $L_c$, depending on $\Delta k$, is strongly influenced both by the separation distance between the waveguides and by the external magnetic field. Bringing the waveguides closer together in the coupling region would indeed lead to a stronger dipolar interaction and thus to a larger energy splitting between the two modes, resulting in an increase of $\Delta k$. Conversely, it is evident that in the limit of non-interacting waveguides, the two modes must collapse into the single mode supported by an isolated waveguide. Similarly, a change in the external field leads, to a first approximation, to a shift of the dispersion relations reported in Fig. 2b, thereby modifying $\Delta k$ at a given excitation frequency.

The transfer matrix in the waveguides basis can be then expressed as:

$$U_{\mathrm{g}} = e^{i\bar{k}L} \begin{bmatrix} \cos\vartheta & -i\sin\vartheta \\ -i\sin\vartheta & \cos\vartheta \end{bmatrix}$$

In the following discussions, the phase factor $e^{i\bar{k}L}$ can be neglected as it is common to both branches. Indeed, as will be emphasized in the next paragraph, the physical quantity of interest for propagation is not the absolute phase of the two waveguides but rather their relative phase shift.

### Single-input excitation of the directional coupler

In the specific case where only one of the two waveguides is excited (see Fig. 2e), the output vector can be simply derived:

$$Z\begin{bmatrix} 0 \\ 1 \end{bmatrix} = U_{\mathrm{g}}\begin{bmatrix} 0 \\ 1 \end{bmatrix} = \begin{bmatrix} \cos\vartheta & -i\sin\vartheta \\ -i\sin\vartheta & \cos\vartheta \end{bmatrix}\begin{bmatrix} 0 \\ 1 \end{bmatrix} = \begin{bmatrix} -i\sin\vartheta \\ \cos\vartheta \end{bmatrix}$$

It is evident that, for this specific configuration, the output waves from the two waveguides are always in phase quadrature, as experimentally shown in Fig. 2g-k. The output powers in the two waveguides can be represented by considering the squared moduli of the two components of the output vector:

$$\begin{cases} P_{out1} \propto \sin^2\vartheta \\ P_{out2} \propto \cos^2\vartheta \end{cases}$$

The ability to control $\vartheta$ via the external magnetic field makes the device with single-input excitation arbitrarily reconfigurable, allowing any combination of output powers between the two branches to be obtained[6].

### Phase shifter

Since the phase shifter is physically localized on one of the two waveguides, it is natural to describe it directly in the waveguide basis. Assuming the phase shifter is placed on the first waveguide (i.e., the one identified by the first component of each column vector), one obtains:

$$S = \begin{bmatrix} e^{i\varphi} & 0 \\ 0 & 1 \end{bmatrix}$$

where the term $e^{i\varphi}$ represents the phase shift introduced by the phase shifter in the first waveguide relative to the second. It is important to recall that this phase shift depends not only on the geometry of the phase shifter but also on the externally applied magnetic field $\mu_0 H$, thereby making the device behavior externally controllable (see Fig. 3a-c).

**Phase-controlled 2 × 2 router**

As shown in Fig. 3d, the phase-controlled 2 x 2 router consists of a phase shifter localized on one of the two waveguides cascaded with a directional coupler. The output vector from the phase shifter therefore represents the input vector of the directional coupler. For this reason, given the matrices of the two constituent elements, the matrix representation $Z$ of the tunable directional coupler can be readily obtained as:

$$Z = U_g \cdot S = \begin{bmatrix} e^{i\varphi} \cos \vartheta & -i \sin \vartheta \\ -i\, e^{i\varphi} \sin \vartheta & \cos \vartheta \end{bmatrix}$$

The more general case in which the stripline simultaneously excites both waveguides is now considered. Since the stripline is orthogonal to both their axes, they are excited in phase. The input vector to the tunable directional coupler is therefore the symmetric mode:

$$v_{in} = \begin{bmatrix} 1 \\ 1 \end{bmatrix}$$

The representative matrix acts on the input vector as reported in the following:

$$Z \begin{bmatrix} 1 \\ 1 \end{bmatrix} = U_\mathrm{g} * S \begin{bmatrix} 1 \\ 1 \end{bmatrix} = \begin{bmatrix} e^{i\varphi} \cos \vartheta & -i \sin \vartheta \\ -i\, e^{i\varphi} \sin \vartheta & \cos \vartheta \end{bmatrix} \begin{bmatrix} 1 \\ 1 \end{bmatrix} = \begin{bmatrix} e^{i\varphi} \cos \vartheta - i \sin \vartheta \\ -i\, e^{i\varphi} \sin \vartheta + \cos \vartheta \end{bmatrix}$$

For $\varphi = 0$, the output will always coincide with the symmetric mode for any value of $\vartheta$. In this case, the power redistribution between the two branches is suppressed. To calculate the output powers, it is convenient to rewrite the output vector by explicitly expressing, for each component, the real and imaginary parts:

$$v_{out} = \begin{bmatrix} e^{i\varphi} \cos \vartheta - i \sin \vartheta \\ -i\, e^{i\varphi} \sin \vartheta + \cos \vartheta \end{bmatrix} = \begin{bmatrix} [\cos \varphi \cos \vartheta] + i[\sin \varphi \cos \vartheta - \sin \vartheta] \\ [\cos \vartheta + \sin \varphi \sin \vartheta] - i[\cos \varphi \sin \vartheta] \end{bmatrix}$$

As done in the previous cases, the squared modulus of the components of the output vector must be considered for the calculation of the two output powers:

$$P_{out1} = (\cos \varphi \cos \vartheta)^2 + (\sin \varphi \cos \vartheta - \sin \vartheta)^2 = 1 - \sin \varphi \sin 2\vartheta$$

$$P_{out2} = (\cos \vartheta + \sin \varphi \sin \vartheta)^2 + (-\cos \varphi \sin \vartheta)^2 = 1 + \sin \varphi \sin 2\vartheta$$

From the result just derived, the fundamental role of the input phase delay between the waveguides introduced by the phase shifter emerges. The term $\sin \varphi$ determines the minimum and maximum achievable power values in each waveguide. Complete energy transfer is possible only when $\varphi = \pm\frac{\pi}{2}$, i.e., under the conditions $\sin \varphi = \pm 1$. At the same time, for a given value of $\vartheta$, changing the sign of $\varphi$ leads to an exchange of the output powers between the two waveguides. Finally, total power conservation is verified, since $P_{out1} + P_{out2} = 2$. The choice not to normalize the total power to 1 stems from the intention to keep track of the double excitation of the waveguides.

To complete the analysis of the output vector, the phases of the individual waveguides must be taken into account,

$$\phi_\mathrm{out1} = \mathrm{atan}\frac{\mathfrak{I}_1}{\mathfrak{R}_1} = \mathrm{atan}\left(\frac{\sin \varphi \cos \vartheta - \sin \vartheta}{\cos \varphi \cos \vartheta}\right)$$

$$\phi_{out2} = \mathrm{atan}\frac{\mathfrak{I}_2}{\mathfrak{R}_2} = \mathrm{atan}\left(\frac{-\cos \varphi \sin \vartheta}{\cos \vartheta + \sin \varphi \sin \vartheta}\right)$$

The relative output phase delay of the first waveguide with respect to the second (in full analogy with the definition of the input phase) can be expressed as the difference between these two phases:

$$\Delta\phi_{\text{out}} = \phi_{\text{out1}} - \phi_{\text{out2}} = \text{atan}\left(\frac{\sin\varphi\cos\vartheta - \sin\vartheta}{\cos\varphi\cos\vartheta}\right) - \text{atan}\left(\frac{-\cos\varphi\sin\vartheta}{\cos\vartheta + \sin\varphi\sin\vartheta}\right) = \text{atan(A)} - \text{atan(B)}$$

$$\Delta\phi_{\text{out}} = \text{atan}(A) - \text{atan}(B) = \text{atan2}(A - B,\ 1 + AB) = \text{atan2}(N_1D_2 - N_2D_1,\ D_1D_2 + N_1N_2)$$

Using the arctangent subtraction formula, the output phase difference between the two waveguides can be expressed analytically, where $N_1$ and $D_1$ represent the numerator and denominator of the fraction defining $A$, and $N_2$ and $D_2$ those defining $B$, respectively.

However, the standard arctan function returns an angle only in the interval $(-\pi/2, \pi/2)$, making the conventional formula valid only when the argument falls within this range. To overcome this ambiguity and obtain a universally valid relation, the atan2 function is employed, which returns the angle in the full range $(-\pi, \pi)$. This formulation eliminates any quadrant ambiguity and is valid for all values of the parameters $\varphi$ and $\vartheta$, thus representing the most robust choice for the analytical and numerical study of the device.

Accordingly, $\Delta\phi_{\text{out}}$ is expressed in closed form as:

$$\Delta\phi_{\text{out}} = \text{atan2}(\sin\varphi\cos 2\vartheta,\ \cos\varphi)$$

Within the invertibility domain of the tangent, this expression can be written in a more explicit form:

$$\Delta\phi_{\text{out}} \approx \tan^{-1}\left(\frac{\sin\varphi\cos 2\vartheta}{\cos\varphi}\right) = \tan^{-1}(\tan\varphi\cos 2\vartheta)$$

Although this substitution restricts the domain of validity, it allows us to make some observations that can consequently be extended to the general case: the possible values of $\Delta\phi_{\text{out}}$ lie between $(-|\varphi|)$ e $(+|\varphi|)$. During propagation within the coupled region, the parameter $\vartheta = \frac{\pi L}{2L_c}$ varies with the propagation distance, and the output phase oscillates with the same period as the power, within the interval $[-|\varphi|, +|\varphi|]$. The physical length $L$ of the coupled region fixes the value of $\vartheta$ determining a specific value of the output phase difference within that interval.

**Supplementary Discussion 1:** Comparison between theoretical and experimental parameters of the single-input directional coupler.

Fig. 2f presents the theoretical plot of the output power and output phase delay as a function of the theoretical coupling parameter $\vartheta_{\mathrm{theo}}$ (see Supplementary Note 2 for analytical modelling). Superimposed on this plot are the experimental vertical dashed lines obtained from the measured output powers that are also reported in Fig. 2g-k (see Supplementary Methods 2 for data extraction). From these experimental data, the expected theoretical input parameter $\vartheta_{\mathrm{theo}}$ is retrieved. For each routing configuration, Table S1 compares the theoretical values of $\vartheta_{\mathrm{theo}}$ with the corresponding experimental values $\vartheta_{\mathrm{exp}}$.

| TR-MOKE map | $\vartheta_{theo}$ | $\vartheta_{exp}$ |
| --- | --- | --- |
| Fig. 2g | 5.50 | 5.41 |
| Fig. 2h | 6.28 | 6.24 |
| Fig. 2i | 6.85 | 6.61 |
| Fig. 2j | 7.65 | 6.94 |
| Fig. 2k | 6.28 | 6.06 |

**Table S1. Comparison between the experimental and theoretical value of $\vartheta$.**
Experimental values of the coupling parameter $\vartheta_{\mathrm{exp}}$ together with the corresponding theoretical values (from Fig. 2f), are listed for each routing configuration shown in Fig. 2g–k.

The table shows excellent agreement between $\vartheta_{\mathrm{theo}}$ and $\vartheta_{\mathrm{exp}}$, where an effective extension of the coupling region of 65 µm has been considered. A larger discrepancy between the experimental and expected theoretical values is observed in Fig. 2j. This can be easily explained by noticing how, only for Fig. 2j, the coupling between the waveguides starts earlier than in the other panels. By increasing, only for this panel, the extension of the coupling region from 65 µm to 71 µm, the experimental value would return to coincide with the theoretical one.

**Supplementary Discussion 2:** Comparison between theoretical and experimental parameters of the phase-controlled 2 x 2 router.

Fig. 3e presents the theoretical plot of the output power and phase delay as a function of the theoretical coupling parameter $\vartheta_{\mathrm{theo}}$ and the theoretical input phase $\varphi_{\mathrm{theo}}$ (see Supplementary Note 2 for analytical modelling). Superimposed on this plot are the experimental points (star markers), obtained from the measured output powers and the output phase delay $\Delta\phi_{\mathrm{out,\,exp}}$ of the directional coupler (see Supplementary Methods 2 for data extraction). From these experimental data, the associated theoretical input parameter $\vartheta_{\mathrm{theo}}$ and $\varphi_{\mathrm{theo}}$ can be retrieved. For each routing configuration, Table S2 compares the theoretical values with the corresponding experimental values $\vartheta_{\mathrm{exp}}$ and $\varphi_{\mathrm{exp}}$.

| Device | TR-MOKE map | $\vartheta_{theo}$ | $\varphi_{theo}$ | $\vartheta_{exp}$ | $\varphi_{exp}$ |
|---|---|---|---|---|---|
| 2 | Fig. 3f | 5.04 | 1.66 | 5.00 | 1.38 |
| | Fig. 3g | 6.11 | 2.97 | - | 2.73 |
| | Fig. 3h | 6.73 | 1.64 | 6.81 | 1.78 |
| 1 | Fig. 3i | 5.21 | -2.01 | 5.07 | -2.51 |
| | Fig. 3j | 6.55 | -1.43 | 6.57 | -1.99 |

**Table S2. Comparison between the experimental and theoretical value of $\vartheta$ and $\varphi$.**
Experimental values of the coupling parameter $\vartheta_{\mathrm{exp}}$ and input phase $\varphi_{\mathrm{exp}}$, together with the corresponding theoretical values (from Fig. 3e), are listed for each routing configuration shown in Fig. 3f–j.

The table shows excellent agreement between $\vartheta_{\mathrm{theo}}$ and $\vartheta_{\mathrm{exp}}$, where an effective extension of the coupling region of 65 μm has been considered. The agreement between the $\vartheta$ values is noticeably better than that between the $\varphi$ values. This apparent issue is solely due to the extent of the region over which the fit is performed: whereas for $\vartheta$ the region of interest spans the entire coupling length, estimating the input phase necessarily requires restricting the fit region to the area immediately preceding the coupling region. A fit performed over such a limited region inevitably yields less accurate estimates. Nevertheless, the agreement between $\varphi_{\mathrm{theo}}$ and $\varphi_{\mathrm{exp}}$ remains excellent and clearly confirms that the sign of the input phase directly affects the output power splitting, as shown in Supplementary Note 2.